\RequirePackage{lineno}
\documentclass[a4paper,11pt]{article}
\usepackage{jheppub} 

\usepackage[T1]{fontenc} 
\usepackage{graphicx,color,dcolumn,booktabs,bm}
\usepackage{longtable,lscape}
\usepackage{amsmath}
\usepackage{indentfirst}
\usepackage{epsfig}
\usepackage{epstopdf}   
\usepackage{slashed}  
\usepackage{color}
\usepackage{multirow}
\usepackage{graphicx,color,dcolumn,booktabs,bm}
\usepackage{verbatim}

\newcommand{\BR}{{\cal B}}

\newcommand{\beq}{\begin{equation}}
\newcommand{\eeq}{\end{equation}}
\newcommand{\bitm}{\begin{itemize}}
\newcommand{\eitm}{\end{itemize}}

\newcommand{\xic}{\Xi^0_c}
\renewcommand\tabcolsep{2.0pt}

\title{\boldmath Measurements of branching fractions and asymmetry parameters of $\Xi^0_c\to \Lambda\bar K^{*0}$, $\Xi^0_c\to \Sigma^0\bar K^{*0}$, and $\Xi^0_c\to \Sigma^+K^{*-}$ decays at Belle}

\newcounter{AffiliationCounter}
\stepcounter{AffiliationCounter}\edef\instBilbao{\protect\theAffiliationCounter}
\stepcounter{AffiliationCounter}\edef\instBonn{\protect\theAffiliationCounter}
\stepcounter{AffiliationCounter}\edef\instBNL{\protect\theAffiliationCounter}
\stepcounter{AffiliationCounter}\edef\instBINP{\protect\theAffiliationCounter}
\stepcounter{AffiliationCounter}\edef\instCharles{\protect\theAffiliationCounter}
\stepcounter{AffiliationCounter}\edef\instChonnam{\protect\theAffiliationCounter}
\stepcounter{AffiliationCounter}\edef\instCincinnati{\protect\theAffiliationCounter}
\stepcounter{AffiliationCounter}\edef\instDESY{\protect\theAffiliationCounter}
\stepcounter{AffiliationCounter}\edef\instFlorida{\protect\theAffiliationCounter}
\stepcounter{AffiliationCounter}\edef\instFuJen{\protect\theAffiliationCounter}
\stepcounter{AffiliationCounter}\edef\instFudan{\protect\theAffiliationCounter}
\stepcounter{AffiliationCounter}\edef\instGiessen{\protect\theAffiliationCounter}
\stepcounter{AffiliationCounter}\edef\instGifu{\protect\theAffiliationCounter}
\stepcounter{AffiliationCounter}\edef\instSokendai{\protect\theAffiliationCounter}
\stepcounter{AffiliationCounter}\edef\instGyeongsang{\protect\theAffiliationCounter}
\stepcounter{AffiliationCounter}\edef\instHanyang{\protect\theAffiliationCounter}
\stepcounter{AffiliationCounter}\edef\instHawaii{\protect\theAffiliationCounter}
\stepcounter{AffiliationCounter}\edef\instKEK{\protect\theAffiliationCounter}
\stepcounter{AffiliationCounter}\edef\instJPARC{\protect\theAffiliationCounter}
\stepcounter{AffiliationCounter}\edef\instHSE{\protect\theAffiliationCounter}
\stepcounter{AffiliationCounter}\edef\instJuelich{\protect\theAffiliationCounter}
\stepcounter{AffiliationCounter}\edef\instIKER{\protect\theAffiliationCounter}
\stepcounter{AffiliationCounter}\edef\instIISERM{\protect\theAffiliationCounter}
\stepcounter{AffiliationCounter}\edef\instIITB{\protect\theAffiliationCounter}
\stepcounter{AffiliationCounter}\edef\instIITH{\protect\theAffiliationCounter}
\stepcounter{AffiliationCounter}\edef\instIITM{\protect\theAffiliationCounter}
\stepcounter{AffiliationCounter}\edef\instIndiana{\protect\theAffiliationCounter}
\stepcounter{AffiliationCounter}\edef\instIHEP{\protect\theAffiliationCounter}
\stepcounter{AffiliationCounter}\edef\instProtvino{\protect\theAffiliationCounter}
\stepcounter{AffiliationCounter}\edef\instVienna{\protect\theAffiliationCounter}
\stepcounter{AffiliationCounter}\edef\instNapoli{\protect\theAffiliationCounter}
\stepcounter{AffiliationCounter}\edef\instTorino{\protect\theAffiliationCounter}
\stepcounter{AffiliationCounter}\edef\instJAEA{\protect\theAffiliationCounter}
\stepcounter{AffiliationCounter}\edef\instJSI{\protect\theAffiliationCounter}
\stepcounter{AffiliationCounter}\edef\instKarlsruhe{\protect\theAffiliationCounter}
\stepcounter{AffiliationCounter}\edef\instKAU{\protect\theAffiliationCounter}
\stepcounter{AffiliationCounter}\edef\instKitasato{\protect\theAffiliationCounter}
\stepcounter{AffiliationCounter}\edef\instKISTI{\protect\theAffiliationCounter}
\stepcounter{AffiliationCounter}\edef\instKorea{\protect\theAffiliationCounter}
\stepcounter{AffiliationCounter}\edef\instKyotoSangyo{\protect\theAffiliationCounter}
\stepcounter{AffiliationCounter}\edef\instKyungpook{\protect\theAffiliationCounter}
\stepcounter{AffiliationCounter}\edef\instLAL{\protect\theAffiliationCounter}
\stepcounter{AffiliationCounter}\edef\instLebedev{\protect\theAffiliationCounter}
\stepcounter{AffiliationCounter}\edef\instLjubljana{\protect\theAffiliationCounter}
\stepcounter{AffiliationCounter}\edef\instLMU{\protect\theAffiliationCounter}
\stepcounter{AffiliationCounter}\edef\instMNIT{\protect\theAffiliationCounter}
\stepcounter{AffiliationCounter}\edef\instMaribor{\protect\theAffiliationCounter}
\stepcounter{AffiliationCounter}\edef\instMPI{\protect\theAffiliationCounter}
\stepcounter{AffiliationCounter}\edef\instMelbourne{\protect\theAffiliationCounter}
\stepcounter{AffiliationCounter}\edef\instMississippi{\protect\theAffiliationCounter}
\stepcounter{AffiliationCounter}\edef\instMiyazaki{\protect\theAffiliationCounter}
\stepcounter{AffiliationCounter}\edef\instMEPhI{\protect\theAffiliationCounter}
\stepcounter{AffiliationCounter}\edef\instNagoya{\protect\theAffiliationCounter}
\stepcounter{AffiliationCounter}\edef\instNagoyaKMI{\protect\theAffiliationCounter}
\stepcounter{AffiliationCounter}\edef\instUNapoli{\protect\theAffiliationCounter}
\stepcounter{AffiliationCounter}\edef\instNara{\protect\theAffiliationCounter}
\stepcounter{AffiliationCounter}\edef\instNCU{\protect\theAffiliationCounter}
\stepcounter{AffiliationCounter}\edef\instNUU{\protect\theAffiliationCounter}
\stepcounter{AffiliationCounter}\edef\instTaiwan{\protect\theAffiliationCounter}
\stepcounter{AffiliationCounter}\edef\instKrakow{\protect\theAffiliationCounter}
\stepcounter{AffiliationCounter}\edef\instNihonDental{\protect\theAffiliationCounter}
\stepcounter{AffiliationCounter}\edef\instNiigata{\protect\theAffiliationCounter}
\stepcounter{AffiliationCounter}\edef\instNovosibirsk{\protect\theAffiliationCounter}
\stepcounter{AffiliationCounter}\edef\instOsakaCity{\protect\theAffiliationCounter}
\stepcounter{AffiliationCounter}\edef\instPNNL{\protect\theAffiliationCounter}
\stepcounter{AffiliationCounter}\edef\instPanjab{\protect\theAffiliationCounter}
\stepcounter{AffiliationCounter}\edef\instPeking{\protect\theAffiliationCounter}
\stepcounter{AffiliationCounter}\edef\instPittsburgh{\protect\theAffiliationCounter}
\stepcounter{AffiliationCounter}\edef\instNPC{\protect\theAffiliationCounter}
\stepcounter{AffiliationCounter}\edef\instRIKENMSL{\protect\theAffiliationCounter}
\stepcounter{AffiliationCounter}\edef\instUSTC{\protect\theAffiliationCounter}
\stepcounter{AffiliationCounter}\edef\instSeoul{\protect\theAffiliationCounter}
\stepcounter{AffiliationCounter}\edef\instShoyaku{\protect\theAffiliationCounter}
\stepcounter{AffiliationCounter}\edef\instSoongsil{\protect\theAffiliationCounter}
\stepcounter{AffiliationCounter}\edef\instSungkyunkwan{\protect\theAffiliationCounter}
\stepcounter{AffiliationCounter}\edef\instSydney{\protect\theAffiliationCounter}
\stepcounter{AffiliationCounter}\edef\instTabuk{\protect\theAffiliationCounter}
\stepcounter{AffiliationCounter}\edef\instTata{\protect\theAffiliationCounter}
\stepcounter{AffiliationCounter}\edef\instTUM{\protect\theAffiliationCounter}
\stepcounter{AffiliationCounter}\edef\instTelAviv{\protect\theAffiliationCounter}
\stepcounter{AffiliationCounter}\edef\instToho{\protect\theAffiliationCounter}
\stepcounter{AffiliationCounter}\edef\instERI{\protect\theAffiliationCounter}
\stepcounter{AffiliationCounter}\edef\instTokyo{\protect\theAffiliationCounter}
\stepcounter{AffiliationCounter}\edef\instTMU{\protect\theAffiliationCounter}
\stepcounter{AffiliationCounter}\edef\instVPI{\protect\theAffiliationCounter}
\stepcounter{AffiliationCounter}\edef\instWayneState{\protect\theAffiliationCounter}
\stepcounter{AffiliationCounter}\edef\instYamagata{\protect\theAffiliationCounter}
\stepcounter{AffiliationCounter}\edef\instYonsei{\protect\theAffiliationCounter}

\collaboration{The Belle Collaboration}
\author[\instFudan]{S.~Jia,} 
\author[\instFudan]{S.~S.~Tang,} 
 \author[\instFudan]{C.~P.~Shen,} 
  \author[\instKEK,\instSokendai]{I.~Adachi,} 
  \author[\instTokyo]{H.~Aihara,} 
  \author[\instTabuk,\instKAU]{S.~Al~Said,} 
  \author[\instBNL]{D.~M.~Asner,} 
  \author[\instBINP,\instNovosibirsk]{V.~Aulchenko,} 
  \author[\instHSE]{T.~Aushev,} 
  \author[\instTabuk]{R.~Ayad,} 
% ...

\abstract{Using a data sample of 980 fb$^{-1}$ collected with the Belle detector at the KEKB asymmetric-energy $e^+e^-$ collider, we study the processes of $\Xi^0_c\to \Lambda\bar K^{*0}$, $\Xi^0_c\to \Sigma^0\bar K^{*0}$, and $\Xi^0_c\to \Sigma^+K^{*-}$ for the first time. The relative branching ratios to the normalization mode of $\Xi^0_c\to\Xi^-\pi^+$ are measured to be
$${\cal B}(\Xi^0_c\to \Lambda\bar K^{*0})/{\cal B}(\xic\to \Xi^-\pi^+)=0.18\pm0.02({\rm stat.})\pm0.01({\rm syst.}),$$
$${\cal B}(\Xi^0_c\to \Sigma^0\bar K^{*0})/{\cal B}(\xic\to \Xi^-\pi^+)=0.69\pm0.03({\rm stat.})\pm0.03({\rm syst.}),$$
$${\cal B}(\Xi^0_c\to \Sigma^+K^{*-})/{\cal B}(\xic\to \Xi^-\pi^+)=0.34\pm0.06({\rm stat.})\pm0.02({\rm syst.}),$$
where the uncertainties are statistical and systematic, respectively.
We obtain
$${\cal B}(\Xi^0_c\to \Lambda\bar K^{*0})=(3.3\pm0.3({\rm stat.})\pm0.2({\rm syst.})\pm1.0({\rm ref.}))\times10^{-3},$$
$${\cal B}(\Xi^0_c\to \Sigma^0\bar K^{*0})=(12.4\pm0.5({\rm stat.})\pm0.5({\rm syst.})\pm3.6({\rm ref.}))\times10^{-3},$$
$${\cal B}(\Xi^0_c\to \Sigma^+K^{*-})=(6.1\pm1.0({\rm stat.})\pm0.4({\rm syst.})\pm1.8({\rm ref.}))\times10^{-3},$$ where the uncertainties are statistical, systematic, and from ${\cal B}(\xic \to \Xi^-\pi^+)$, respectively. The asymmetry parameters $\alpha(\Xi^0_c\to \Lambda\bar K^{*0})$ and $\alpha(\Xi^0_c\to \Sigma^+K^{*-})$ are $0.15\pm0.22({\rm stat.})\pm0.04({\rm syst.})$ and $-0.52\pm0.30({\rm stat.})\pm0.02({\rm syst.})$, respectively, where the uncertainties are statistical followed by systematic.}


\begin{document}
\maketitle
\flushbottom

\section{Introduction}

In comparison with the lowest-lying charmed baryon state, the $\Lambda^+_c$, our knowledge of the $\Xi_c$ states is still limited~\cite{PDG}. Recently, there have been many measurements of the lifetime and decay modes of the $\Xi_c$ made by several experiments.
The lifetimes of the $\Xi^0_c$ and $\Xi^+_c$ are $(154.5\pm2.5)$ fs and $(456.8\pm5.5)$ fs~\cite{lifetime}, respectively.
The absolute branching fraction of $\Xi^0_c\to \Xi^-\pi^+$ has been measured to be $(1.80\pm0.52)$\%, so that now the branching fractions of other channels can be determined from ratios of branching fractions~\cite{Li}.
The first branching fraction of the decay of the $\xic$ to a charmed baryon has been measured to be $\BR(\xic\to \Lambda^+_c\pi^-)$ = $(0.55\pm0.02\pm0.18)$\%~\cite{071101}.
The branching fraction ratios of resonant and non-resonant decays of $\Xi^0_c \to \Xi^0 K^+K^-$ with respect to $\Xi^0_c \to \Xi^-\pi^+$ are $0.036\pm0.004\pm0.002$ and $0.039\pm0.004\pm0.002$~\cite{McNeil}, respectively. The branching fractions of semileptonic decays $\Xi^0_c \to \Xi^- e^+\nu_e$ and $\Xi^0_c \to \Xi^- \mu^+\nu_\mu$ have been measured, with much improved precision than hitherto, to be $(1.72\pm0.10\pm0.12\pm0.50)$\% and $(1.71\pm0.17\pm0.13\pm0.50)$\%~\cite{1548}, respectively, where the first, second, and third uncertainties are statistical, systematic, and from $\BR(\xic \to \Xi^-\pi^+)$~\cite{Li}. The decay asymmetry parameter $\alpha(\Xi^0_c \to \Xi^-\pi^+)$ has been measured to be $0.59\pm0.03\pm0.02$~\cite{1548}.

Unlike semileptonic decays of $\Xi^0_c$, which proceed via weak decay processes mediated by $W$ bosons, the non-leptonic weak decays are caused by the $W$-boson exchanges with QCD corrections~\cite{053002}. However, as the strong coupling is large at the typical energies of charm decays, it is very difficult to make quantitative predictions of decay rates and asymmetries with QCD corrections.
Theoretical calculations for the hadronic decays of the $\Xi_c$ have been performed based on $SU(3)_F$ flavor symmetry~\cite{053002,00876,1527,414,7067,2132,147,073006,593,073003,19,214,114022,946,165,066,03480,35,036018} and dynamical models~\cite{659,3836,270,4188,014011,5787,3417,5632,217}.
The two-body ${\bf B}_c \to {\bf B}_n V$ decays have been calculated, where ${\bf B}_c$ and ${\bf B}_n$ correspond to the antitriplet charmed and light baryons, and $V$ stands for the vector mesons, respectively. However, the different models give widely varying predictions.~For $\Xi^0_c\to \Lambda\bar K^{*0}$, $\Xi^0_c\to \Sigma^0\bar K^{*0}$, and $\Xi^0_c\to \Sigma^+K^{*-}$, the predicted branching fractions cover wide ranges of $(0.46 - 1.55)\%$, $(0.27 - 0.85)\%$, and $(0.24 - 0.93)\%$~\cite{659,5787,35,053002}, as listed in Table~\ref{br} for three different models.

\begin{table}[htbp!]
\caption{Decay branching fractions (\%) of the Cabibbo-favored $\xic \to {\bf B}_n V$ decays based on the covariant quark model from KK~\cite{659}, pole model from Zen~\cite{5787}, and $SU(3)_F$ flavor symmetry from HYZ~\cite{35} and GLT~\cite{053002}.}\label{br}
\vspace{0.2cm}
\centering
\begin{tabular}{c c c c c}
\hline\hline
Channel & KK~\cite{659} & Zen~\cite{5787} & HYZ~\cite{35} & GLT~\cite{053002} \\\hline
$\xic \to \Lambda\bar K^{*0}$ & 1.55 & 1.15 & 0.46$\pm$0.21 & 1.37$\pm$0.26 \\
$\xic \to \Sigma^0\bar K^{*0}$ & 0.85 & 0.77 & 0.27$\pm$0.22 & 0.42$\pm$0.23\\
$\xic \to \Sigma^+K^{*-}$ & 0.54 & 0.37 & 0.93$\pm$0.29 & 0.24$\pm$0.17 \\
\hline\hline
\end{tabular}
\end{table}

For the channels $\Xi^0_c\to \Lambda\bar K^{*0}$, $\Xi^0_c\to \Sigma^0\bar K^{*0}$, and $\Xi^0_c\to \Sigma^+K^{*-}$, parity violation is manifested by polarization of the hyperon ($\Lambda$, $\Sigma^0$, or $\Sigma^+$), and is quantified by the decay asymmetry parameter, $\alpha$. Because the hyperon decay also violates parity, the product of decay asymmetry parameters of $\Xi^0_c$ decay and hyperon decay can be measured by its decay angular distribution. Note that the asymmetry parameter of $\Xi^0_c\to \Sigma^0\bar K^{*0}$ can not be measured since the value of $\alpha(\Sigma^0\to\gamma\Lambda)$ should be zero for an electromagnetic decay of $\Sigma^0\to\gamma\Lambda$. We measure the product value of $\alpha(\xic\to\Sigma^0 {\bar K}^{*0})\alpha(\Sigma^0\to \gamma \Lambda)$ just to validate no bias in the measurement.
Different models produce widely different predictions for $\alpha$~\cite{659,5787,053002}, which are listed in Table~\ref{asyt}.

\begin{table}[htbp!]
\caption{The asymmetry parameters for the Cabibbo-favored $\xic \to {\bf B}_n V$ decays based on the covariant quark model from KK~\cite{659}, pole model from Zen~\cite{5787}, and $SU(3)_F$ flavor symmetry from GLT~\cite{053002}.}\label{asyt}
\vspace{0.2cm}
\centering
\begin{tabular}{c c c c c}
\hline\hline
Channel & KK~\cite{659} & Zen~\cite{5787} & GLT~\cite{053002} \\\hline
$\xic \to \Lambda\bar K^{*0}$ & $+0.58$ & $+0.49$ & $-0.67\pm0.24$ \\
$\xic \to \Sigma^0\bar K^{*0}$ & $-0.87$ & $+0.25$ & $-0.42\pm0.62$ \\
$\xic \to \Sigma^+K^{*-}$ & $-0.60$ & $+0.51$ & $-0.76^{+0.64}_{-0.24}$ \\
\hline\hline
\end{tabular}
\end{table}

In this article, we use the entire data sample of 980 fb$^{-1}$ integrated luminosity collected at Belle to perform the first measurements of the branching fractions and asymmetry parameters for the decays $\xic \to \Lambda\bar K^{*0}$, $\xic \to \Sigma^0\bar K^{*0}$, and $\xic \to \Sigma^+K^{*-}$.
The $\Lambda$, $\bar K^{*0}$, $\Sigma^0$, $\Sigma^+$, and $K^{*-}$ are reconstructed by $p\pi^-$, $K^-\pi^+$, $\Lambda\gamma$, $p\pi^0$, and $K^0_S\pi^-$ final states, respectively. Throughout this analysis, for any given mode, the corresponding charge-conjugate mode is implied.

\section{Data sample and the Belle Detector}

This measurement is based on data recorded at or near the $\Upsilon(1S)$, $\Upsilon(2S)$, $\Upsilon(3S)$, $\Upsilon(4S)$, and $\Upsilon(5S)$ resonances by the Belle detector~\cite{Belle1,Belle2} at the KEKB asymmetric-energy $e^+e^-$ collider~\cite{KEKB1,KEKB2}. The total data sample corresponds to an integrated luminosity of 980 fb$^{-1}$~\cite{Belle2}. The Belle detector is a large solid-angle magnetic spectrometer that consists of a silicon vertex detector, a 50-layer central drift chamber (CDC), an array of aerogel threshold Cherenkov counters (ACC), a barrel-like arrangement of time-of-flight scintillation counters (TOF), and an electromagnetic calorimeter comprised of CsI(TI) crystals (ECL) located inside a superconducting solenoid coil that provides a 1.5T magnetic field. An iron flux-return yoke instrumented with resistive plate chambers located outside the coil is used to detect $K^0_L$ mesons and identify muons. A detailed description of the Belle detector can be found in Refs.~\cite{Belle1,Belle2}.

Samples of simulated signal events are generated using {\sc evtgen}~\cite{EVTGEN} to optimize the signal selection criteria and calculate the signal reconstruction efficiency; $e^+e^-\to c\bar c$ events are simulated using {\sc pythia}~\cite{PYTHIA}, and $\xic\to \Lambda \bar K^{*0}/\Sigma^0\bar K^{*0}/\Sigma^+K^{*-}$ decays are generated with a phase space model.
The effect of final-state radiation is taken into account in the simulation using the PHOTOS~\cite{291} package. The simulated events are processed with a detector simulation based on {\sc geant3}~\cite{geant3}. Generic simulated samples, i.e.~$B$ = $B^+$, $B^0$, or $B^{(*)}_s$ decays and $e^+e^-\to q\bar q$ ($q$ = $u$, $d$, $s$, $c$) at $\sqrt{s}$ = 10.52, 10.58, and 10.867 GeV, normalized to the same integrated luminosity as real data, are used to check peaking backgrounds and to perform input/output checks.

\section{Selection criteria}

Except for the charged tracks from the relatively long-lived $\Lambda \to p\pi^-$ and $K^0_S\to \pi^+\pi^-$ decays, impact parameters with respect to the interaction point (IP) are required to be less than 1 cm and 4 cm perpendicular to, and along the beam direction, respectively.~For the particle identification (PID) of a well-reconstructed charged track, information from different detector subsystems, including specific ionization in the CDC, time measurement in the TOF, and the response of the ACC, is combined to form a likelihood ${\cal L}_i$~\cite{PID} for particle species $i$. Tracks with $R_K={\cal L}_K/({\cal L}_K+{\cal L}_\pi)$ $<$ 0.4 are identified as pions with an efficiency of 95\%, while 6\% of kaons are misidentified as pions; tracks with $R_K$ $>$ 0.6 are identified as kaons with an efficiency of 96\%, while 7\% of pions are misidentified as kaons. For proton identification, a track with $R^\pi_p={\cal L}_p/({\cal L}_p+{\cal L}_\pi)$ $>$ 0.6 and $R^K_p={\cal L}_p/({\cal L}_p+{\cal L}_K)$ $>$ 0.6 is identified as a proton with an efficiency of about 98\%; less than 1\% of the pions/kaons are misidentified as protons.

Using a multivariate analysis with a neural network~\cite{190} based on two sets of input variables~\cite{2014}, a $K^0_S$ candidate is reconstructed from a pair of oppositely charged tracks that are treated as pions.~The invariant mass of the $K^0_S$ candidates is required to be within 10 MeV/$c^2$ of the corresponding nominal mass~\cite{PDG}. Candidate $\Lambda$ baryons are reconstructed in the decay $\Lambda\to p\pi^-$ and selected if $|M(p\pi^-)-m_\Lambda|$ $<$ 3 MeV/$c^2$ ($\sim2.5\sigma$), where $\sigma$ denotes the mass resolution. Hereinafter, $M$ represents a measured invariant mass and $m_i$ denotes the nominal mass of particle $i$~\cite{PDG}.

An ECL cluster is treated as a photon candidate if it is isolated from the projected path of charged tracks in the CDC, and its energy in the laboratory frame is greater than 50 MeV. To reduce photon candidates originating from neutral hadrons, we reject a photon candidate if the ratio of energy deposited in the central 3 $\times$ 3 square of cells to that deposited in the enclosing 5 $\times$ 5 square of cells in its ECL cluster is less than 0.85.

For the photons from $\Sigma^0\to\gamma \Lambda$ decays, we require $E_\gamma$ $>$ 0.1 GeV in the laboratory frame to suppress backgrounds. Photon pairs are kept as $\pi^0$ candidates if their invariant mass lies in the range 120 MeV/$c^2$ $<$ $M(\gamma\gamma)$ $<$ 150 MeV/$c^2$ ($\pm3\sigma$ about the nominal mass of $\pi^0$). In the $\Sigma^+ \to p\pi^0$ reconstruction, combinations of $\pi^0$ candidates and protons are made using those protons with a significantly large ($>$ 1mm) impact parameter with respect to the IP.
The flight directions of $\Sigma^+$ candidates, which are reconstructed from their fitted production and decay vertices, are required to be consistent with their momentum directions~\cite{052011}.

The $\Lambda\bar K^{*0}$, $ \Sigma^0\bar K^{*0}$, or $\Sigma^+K^{*-}$ candidates are combined to form a $\xic$ with its daughter tracks fitted to a common vertex. To reduce combinatorial backgrounds, especially from $B$-meson decays, the scaled momentum $x_p = p^*/p_{\rm max}$ is required to
be greater than 0.5. Here, $p^*$ is the momentum of $\xic$ in the center-of-mass (C.M.) frame, and $p_{\rm max}$ = $\sqrt{E^2_{\rm beam}-M^2_{\xic}c^4}/c$
is the maximum momentum, where $E_{\rm beam}$ is the beam energy.

We veto $\Sigma(1385)$ intermediate backgrounds by requiring $M(\Lambda\pi^+)$ $>$ 1.42 GeV/$c^2$ and $M(\Sigma^0\pi^+/\Sigma^+\pi^-)$ $>$ 1.45 GeV/$c^2$ in the substructures of $\Xi^0_c\to \Lambda\bar K^{*0}$ and $\Xi^0_c\to \Sigma^0\bar K^{*0}/\Sigma^+K^{*-}$ candidates, respectively. These requirements can also suppress backgrounds from $D^{*+}$ decays, for which the momentum of pions from $D^{*+}$ is low. No peaking backgrounds are found in $M(\Lambda K^-)$, $M(\Sigma^0 K^-)$, and $M(\Sigma^+ K^0_S)$ distributions from generic simulated samples~\cite{107540}.

To determine the absolute branching fractions of $\Xi^0_c\to \Lambda\bar K^{*0}$, $\Xi^0_c\to \Sigma^0\bar K^{*0}$, and $\Xi^0_c\to \Sigma^+K^{*-}$, the reference mode of $\Xi^0_c\to \Xi^-\pi^+$ is utilized. Selections of candidates in $\Xi^0_c\to \Xi^-\pi^+$ use well-reconstructed tracks, PID, and the vertex fitting technique in a way similar to the methods in Ref.~\cite{1548}. Candidate $\Lambda$ baryons are reconstructed as above. We define the $\Xi^-$ signal region as $|M(\Lambda\pi^-) - m_{\Xi^-}|$ < 6.5 MeV/$c^2$ ($\sim$3$\sigma$). To suppress the combinational background, we require the flight directions of $\Lambda$ and $\Xi^-$ candidates, which are reconstructed from their fitted production and decay vertices, to be within five degrees of their momentum directions. We also require the scaled momentum $x_p$ $>$ 0.5.

\section{Branching fractions of $\Xi^0_c\to \Lambda\bar K^{*0}$, $\Xi^0_c\to \Sigma^0\bar K^{*0}$, and $\Xi^0_c\to \Sigma^+K^{*-}$}

\begin{figure*}[htbp]
\centering
\includegraphics[width=4.8cm,angle=0]{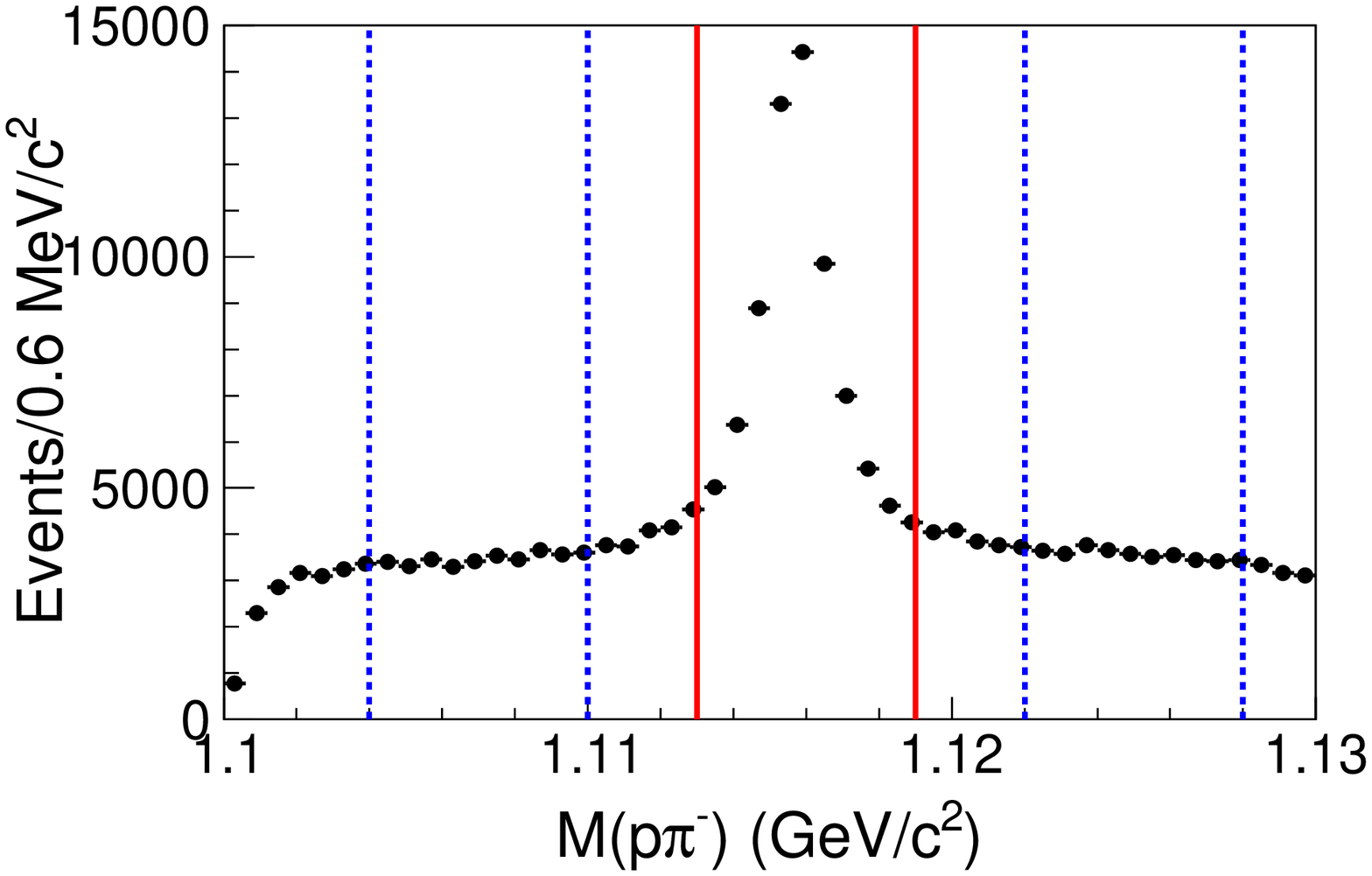}
\put(-32, 70){\bf (a)}
\hspace{0.10cm}
\includegraphics[width=4.8cm,angle=0]{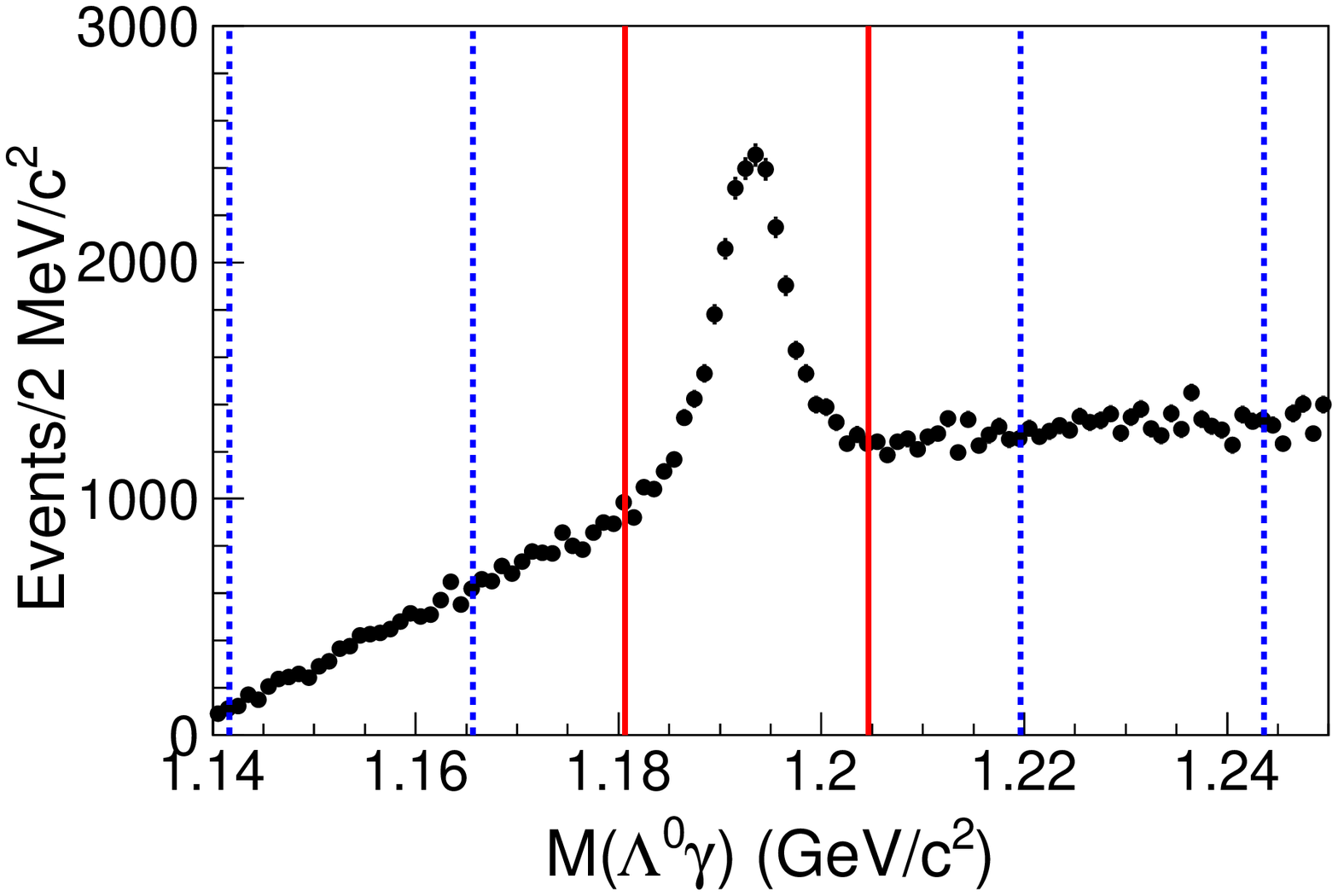}
\put(-32, 70){\bf (b)}
\hspace{0.10cm}
\includegraphics[width=4.8cm,angle=0]{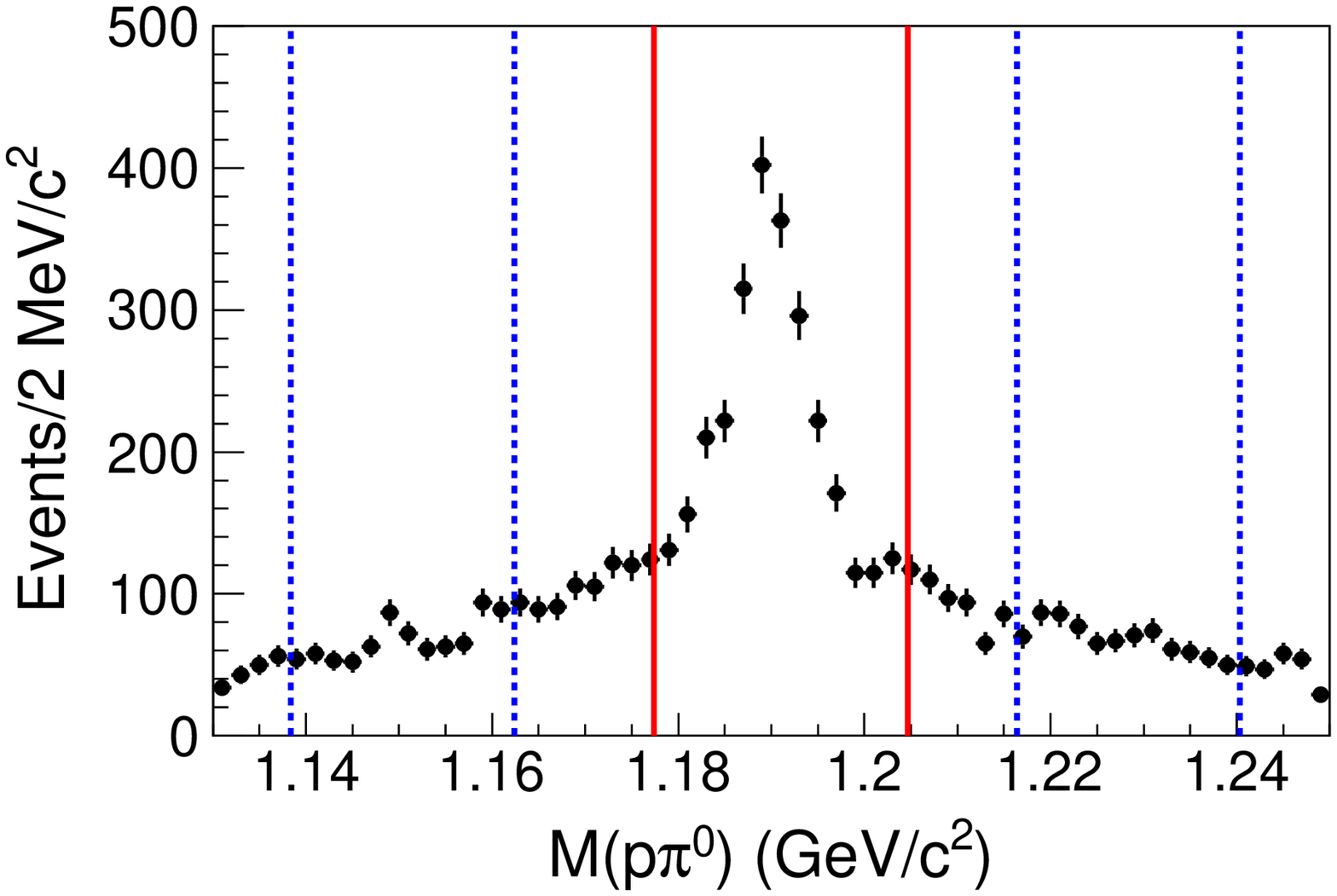}
\put(-32, 70){\bf (c)}
\caption{The invariant mass distributions for $\Lambda$, $\Sigma^0$, and $\Sigma^+$ candidates from data samples. The red solid lines show the required signal regions, and the blue dashed lines show the defined mass sidebands.}\label{data1}
\end{figure*}

After applying the above requirements, the invariant mass distributions of $p\pi^-$, $\Lambda\gamma$, and $p\pi^0$ from data samples are shown in Fig.~\ref{data1}. The $\Lambda$, $\Sigma^0$, and $\Sigma^+$ signals are clear. We define $\Lambda$, $\Sigma^0$, and $\Sigma^+$ signal regions as $|M(p\pi^-) - m_{\Lambda}|$ $<$ 3 MeV/$c^2$ ($\sim2.5\sigma$), $|M(\Lambda\gamma) - m_{\Sigma^0}|$ $<$ 12 MeV/$c^2$ ($\sim2.5\sigma$), and
$|M(p\pi^0) - m_{\Sigma^+}|$ $<$ 12 MeV/$c^2$ ($\sim2.5\sigma$). The mass sideband regions, which will be taken to study backgrounds to our signals, are twice as wide as the signal regions, as shown by the blue dashed lines in Fig.~\ref{data1}.

\begin{figure*}[htbp]
\centering
\includegraphics[width=3.cm,angle=-90]{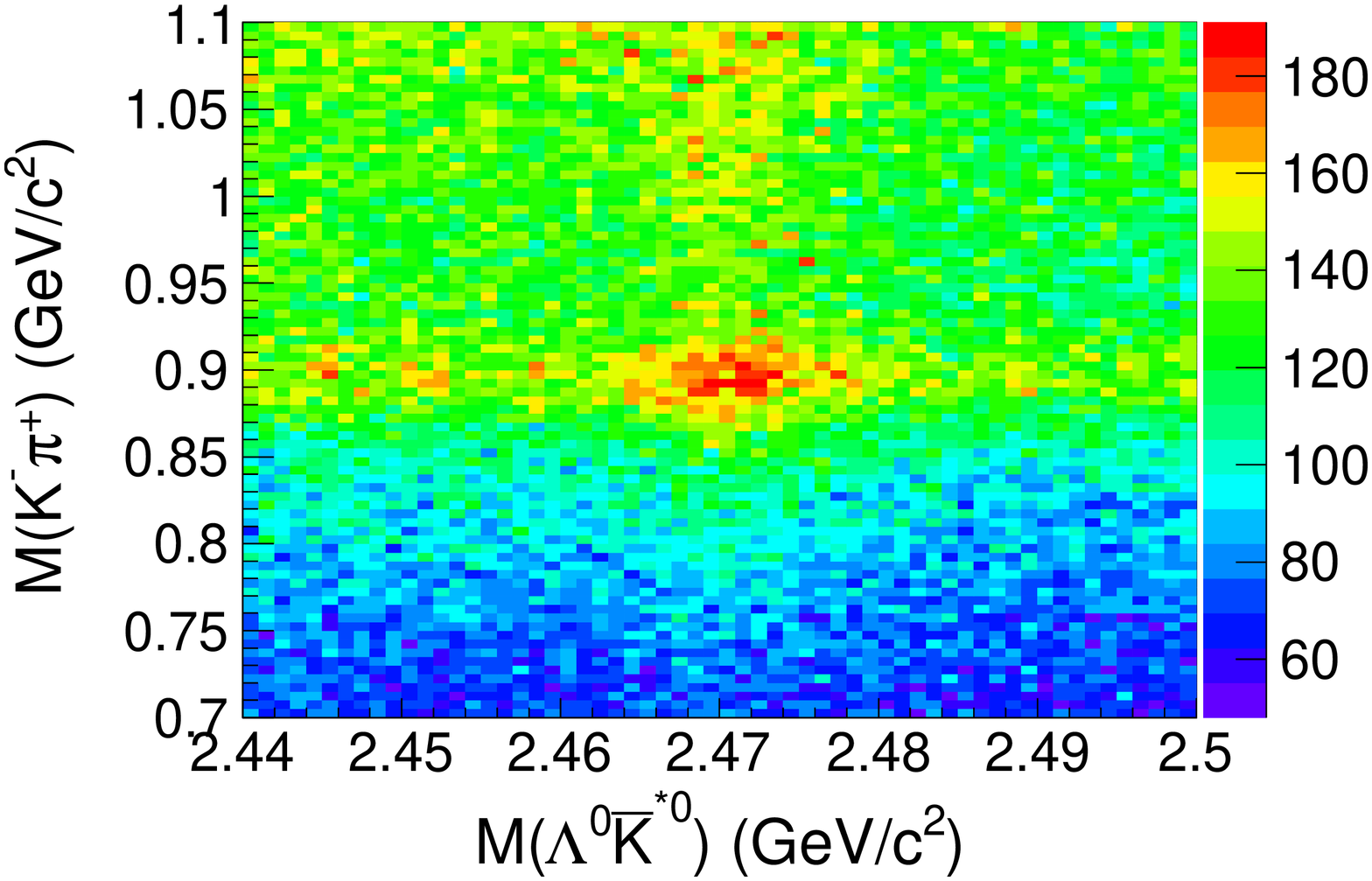}
\put(-134, -5){\scriptsize \bf (a)}
\hspace{0.10cm}
\includegraphics[width=3.cm,angle=-90]{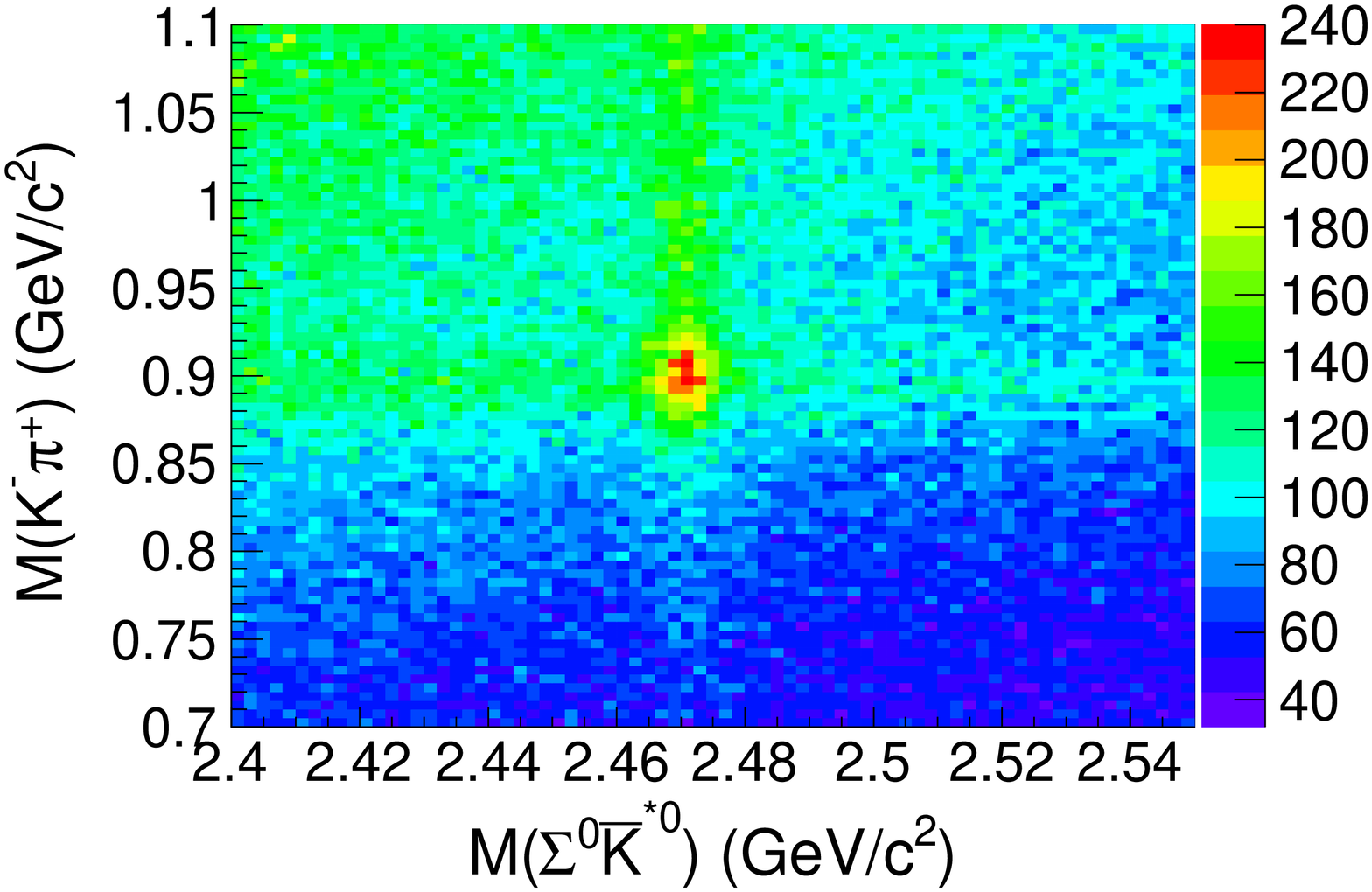}
\put(-134, -5){\scriptsize \bf (b)}
\hspace{0.10cm}
\includegraphics[width=3.cm,angle=-90]{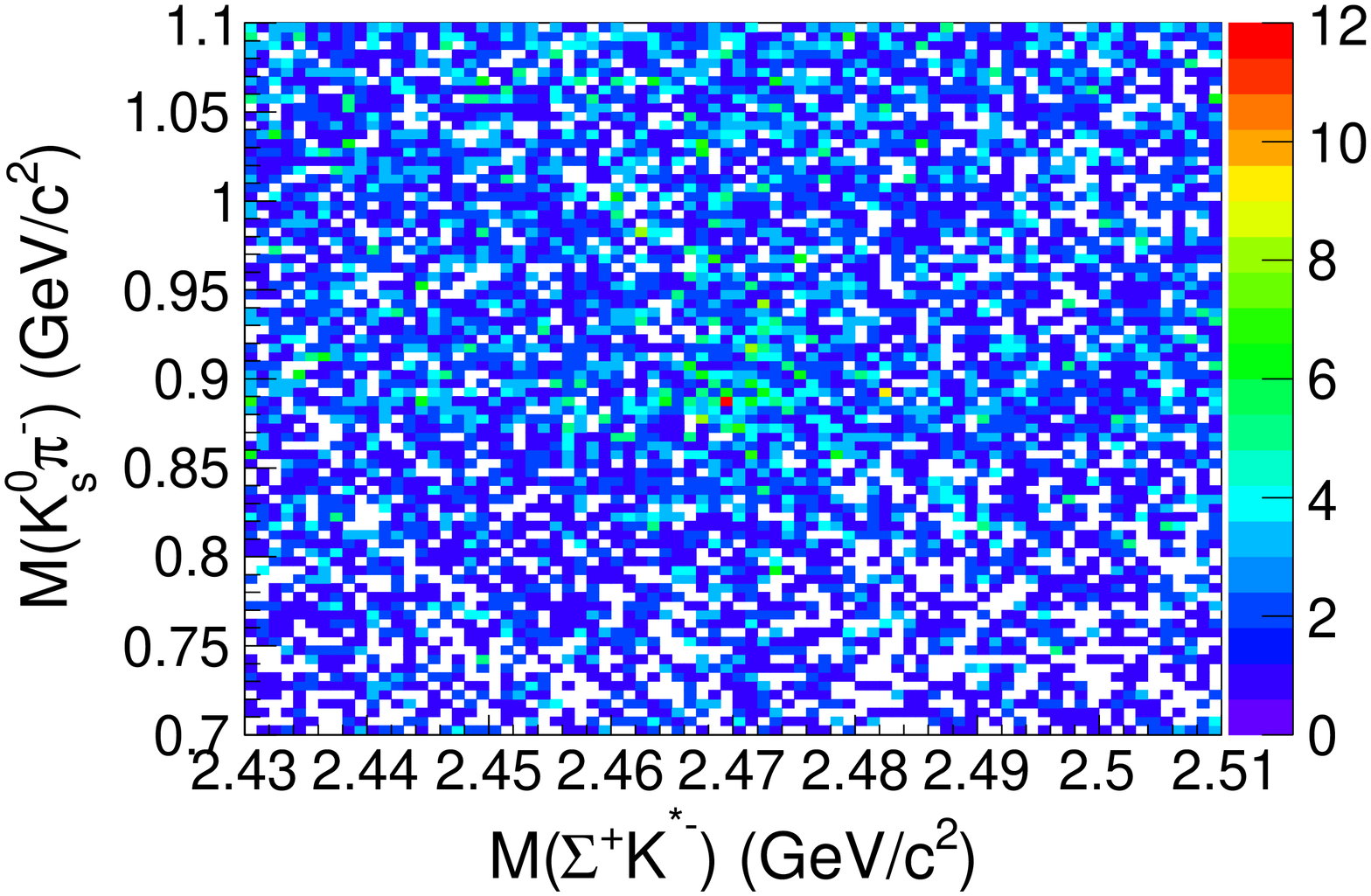}
\put(-134, -5){\scriptsize \bf (c)}
\caption{The scatter plots of (a) $M(K^-\pi^+)$ versus $M(\Lambda\bar K^{*0})$, (b) $M(K^-\pi^+)$ versus $M(\Sigma^0\bar K^{*0})$, and (c) $M(K^0_S\pi^-)$ versus $M(\Sigma^+K^{*-})$ distributions from data samples.}\label{scatter}
\end{figure*}

In the $\Lambda\bar K^{*0}$, $\Sigma^0\bar K^{*0}$, and $\Sigma^+K^{*-}$ invariant mass spectra, no peaking background is found from combinatorial backgrounds of $\Lambda$, $\Sigma^0$, and $\Sigma^+$ candidates, but we found a fraction of $\xic$ signal events from the events outside of the $\bar K^{*0}$ and $K^{*-}$ signal regions, especially in the higher side of $M(K^-\pi^+)$ and $M(K^0_S\pi^-)$, as shown in Fig.~\ref{scatter}.

\begin{figure*}[htbp]
\centering
\includegraphics[width=3.2cm,angle=-90]{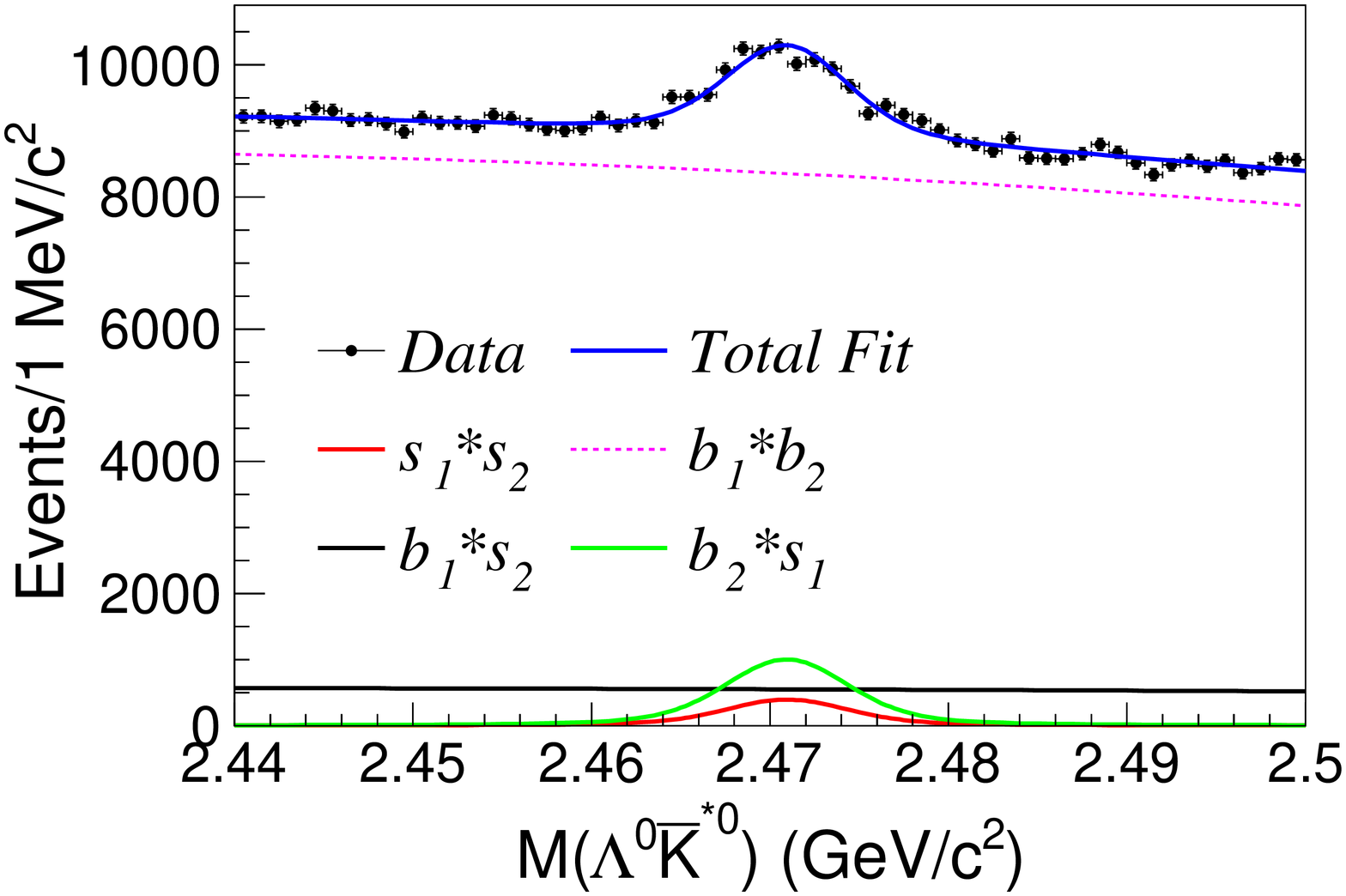}
\includegraphics[width=3.2cm,angle=-90]{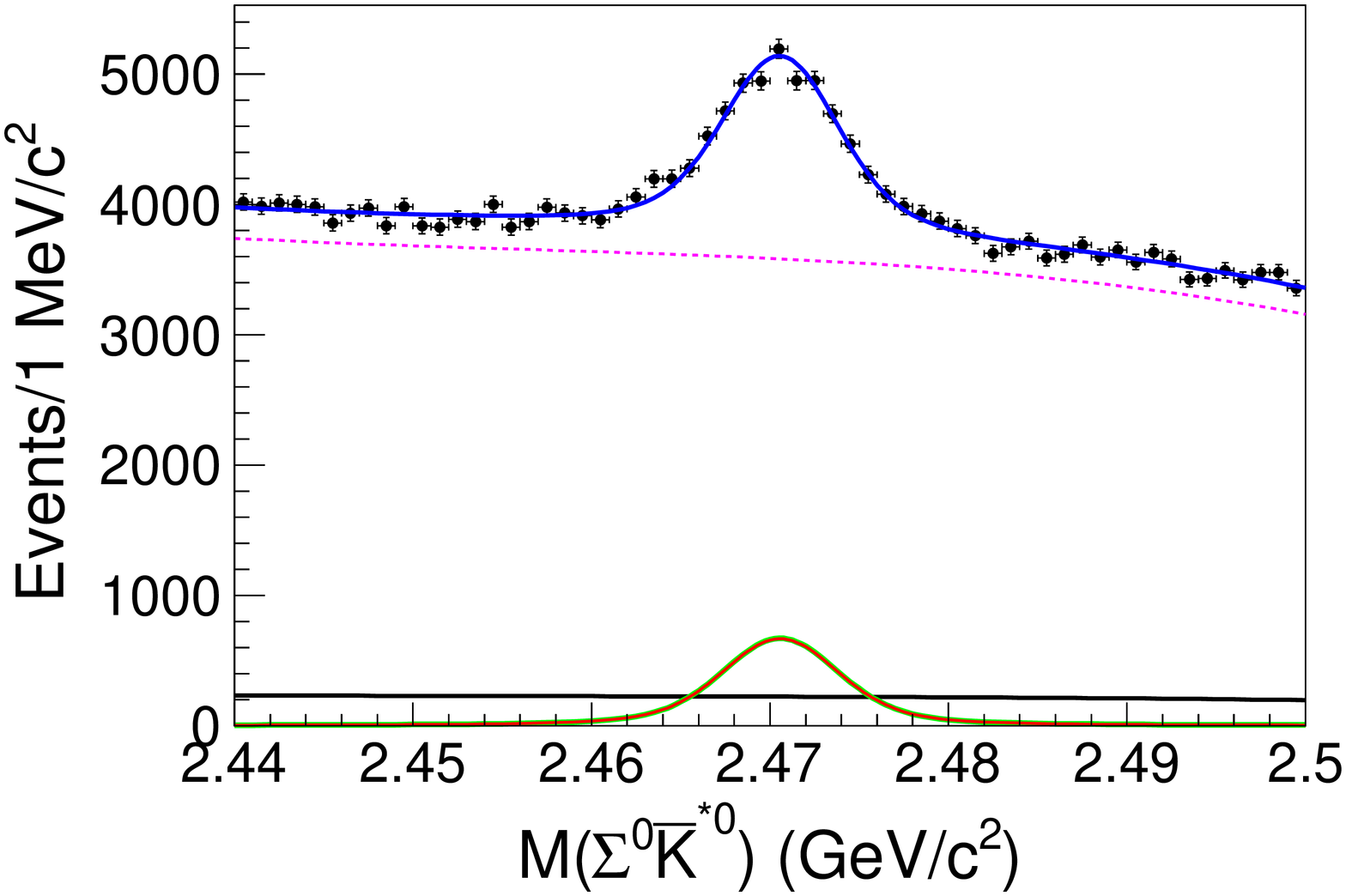}
\includegraphics[width=3.2cm,angle=-90]{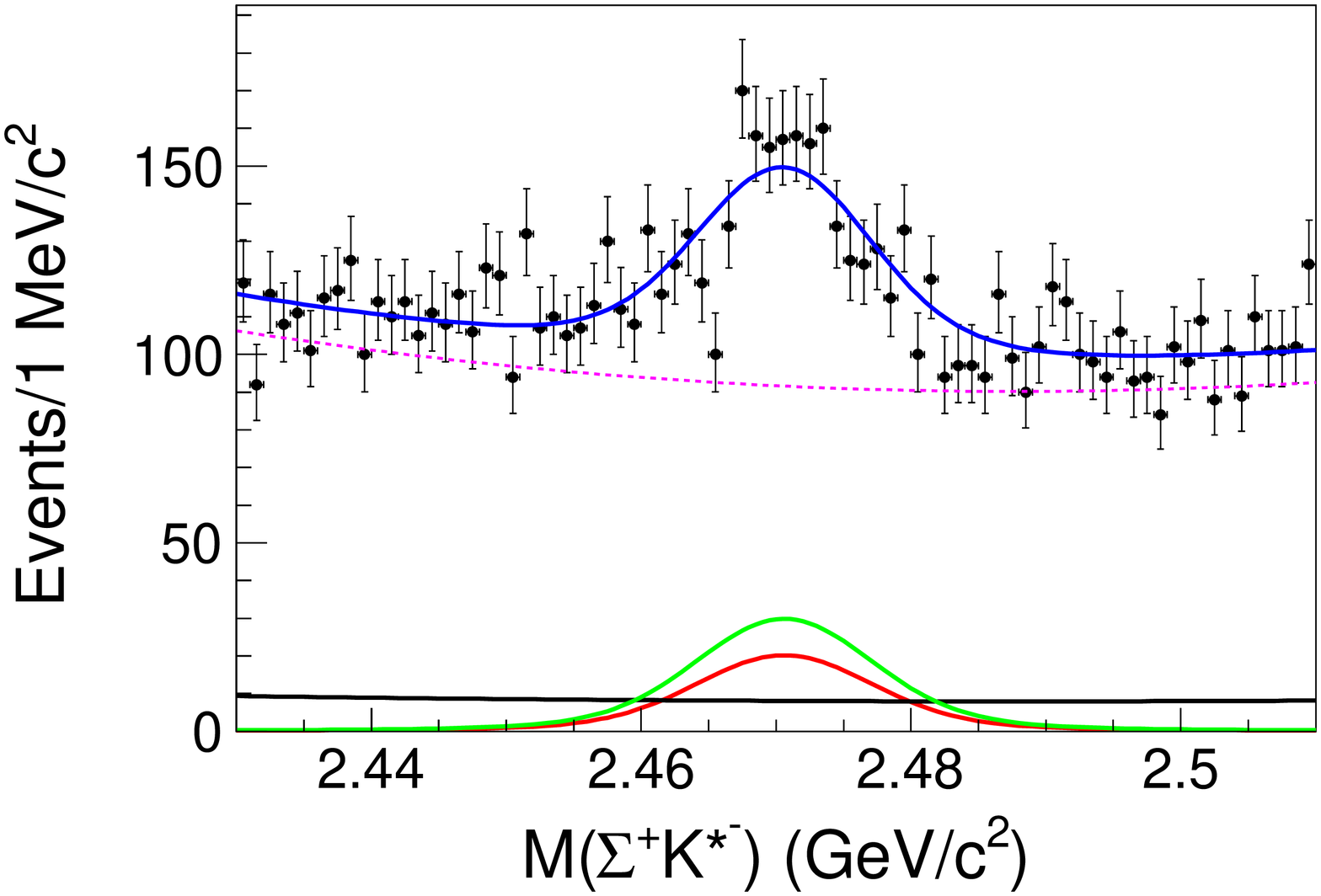}
\vspace{0.3cm}

\includegraphics[width=3.2cm,angle=-90]{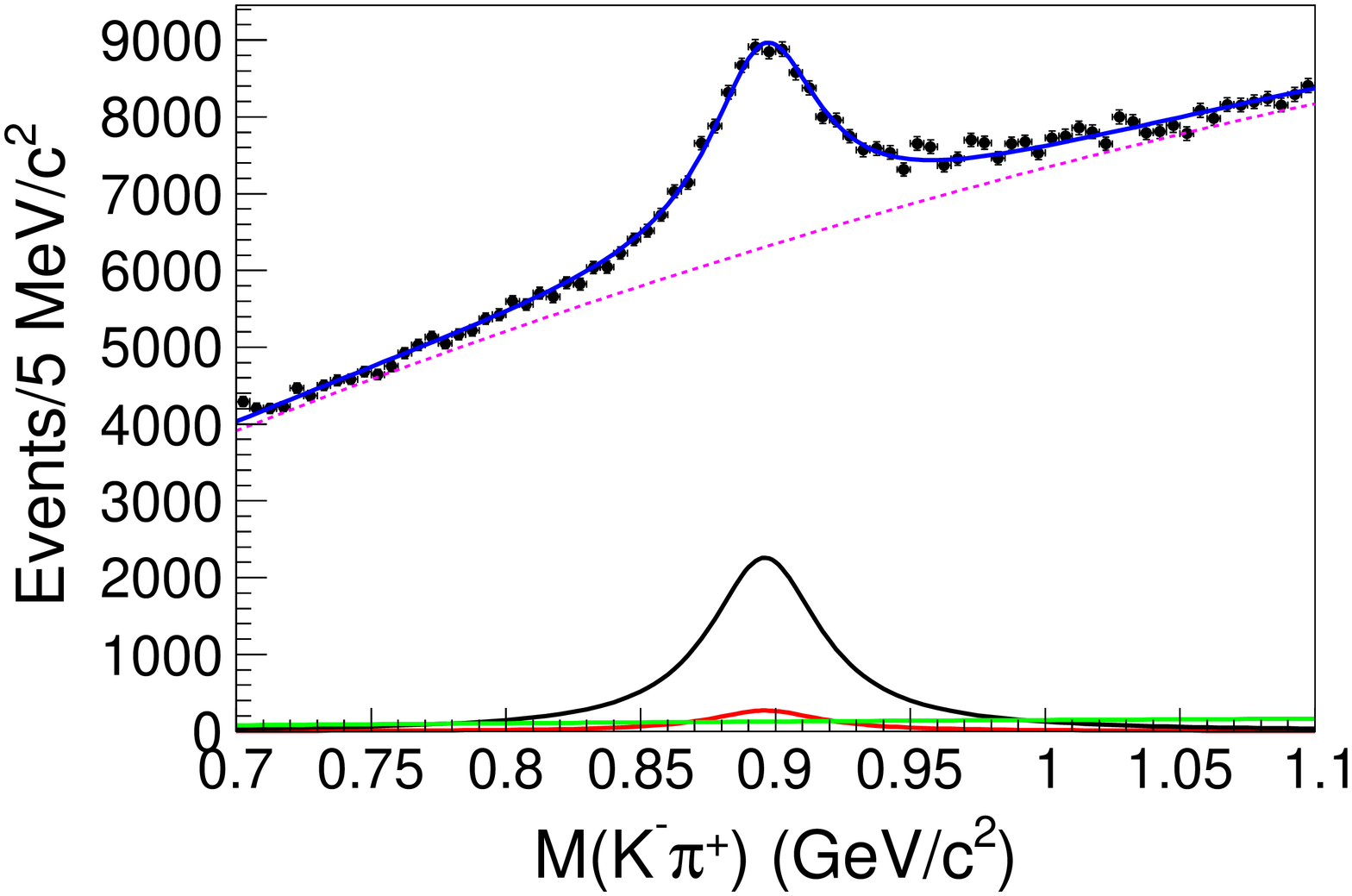}
\includegraphics[width=3.2cm,angle=-90]{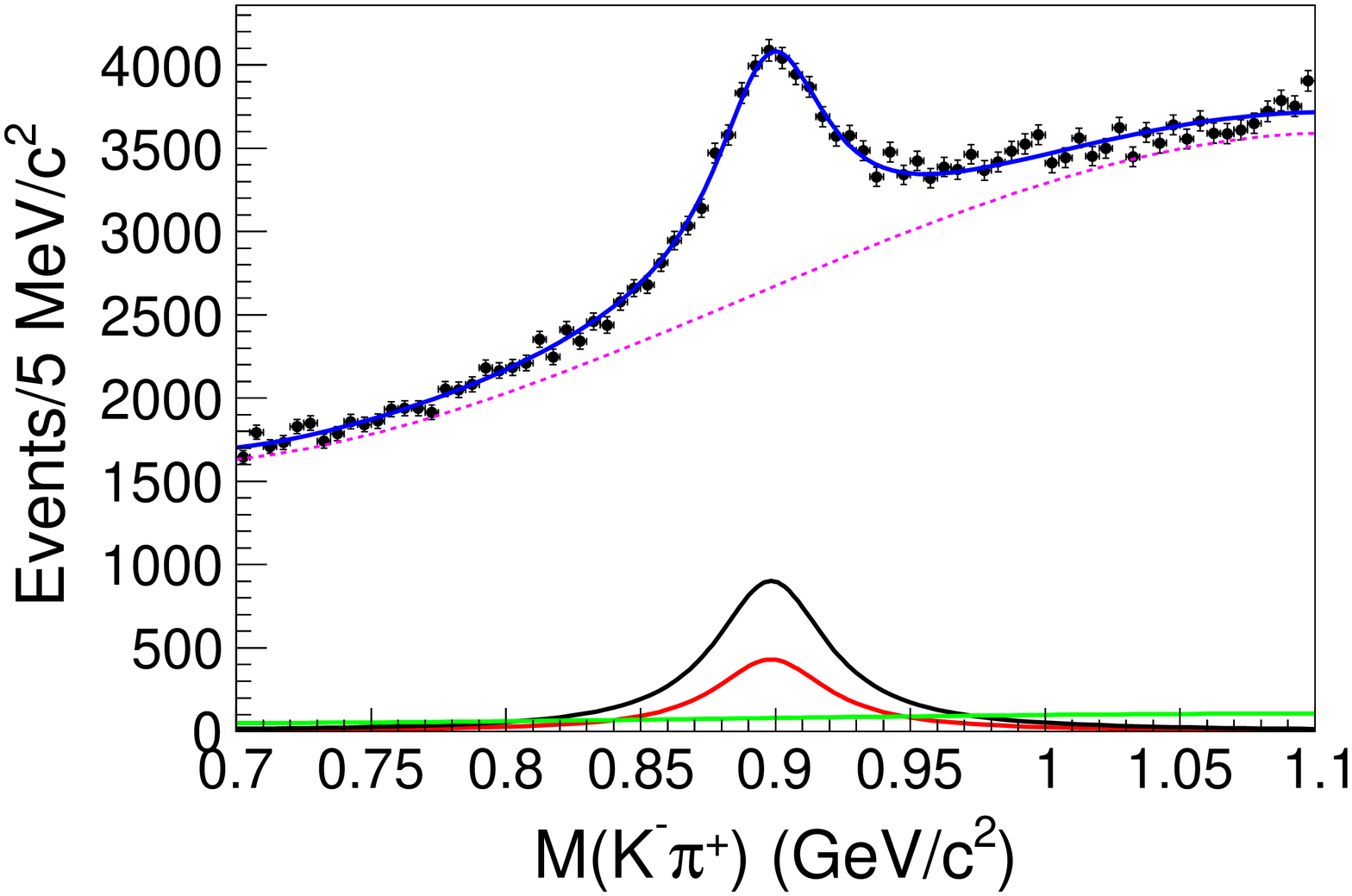}
\includegraphics[width=3.2cm,angle=-90]{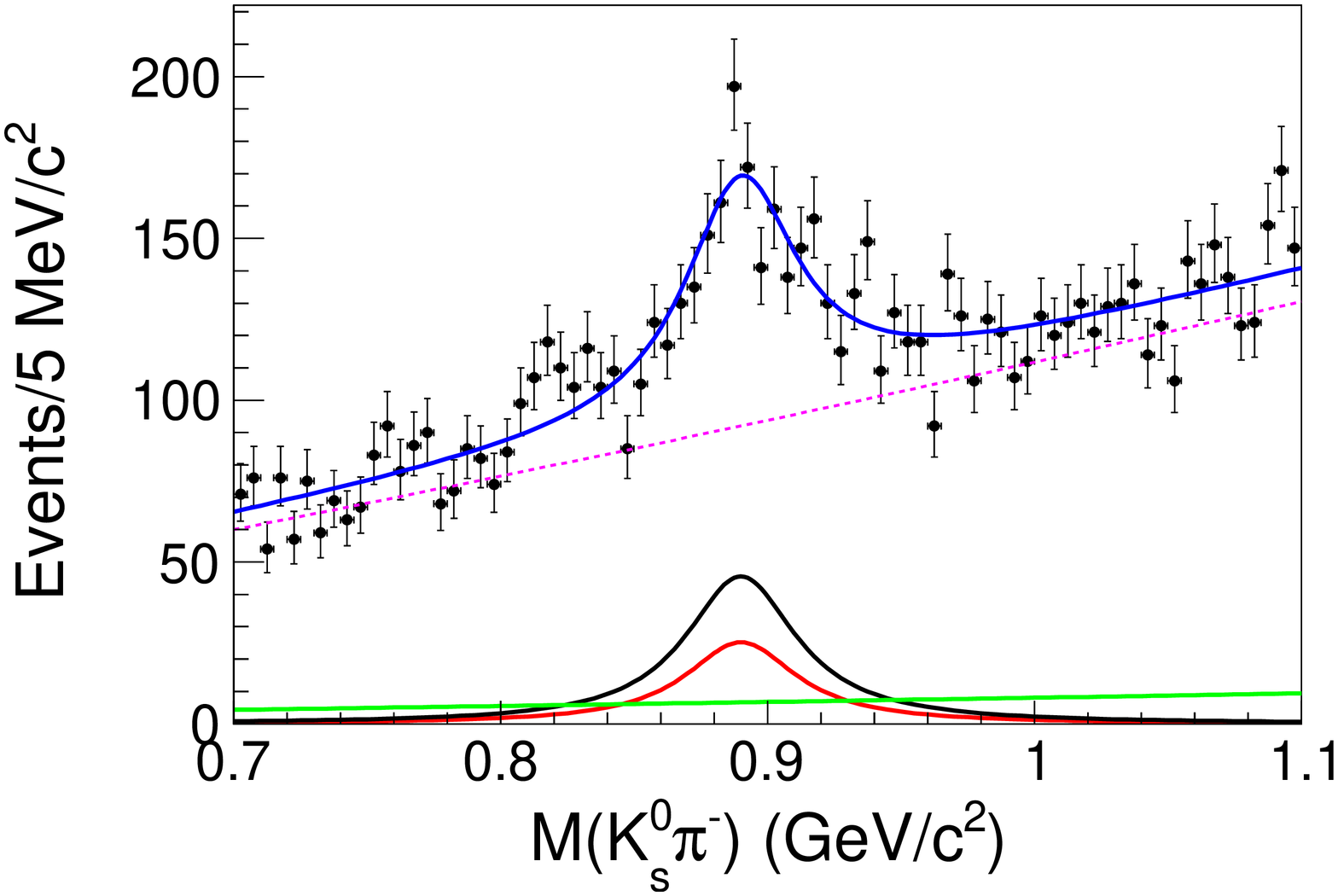}
\caption{2D binned maximum-likelihood fits to $M(K^-\pi^+)$ and $M(\Lambda\bar K^{*0})$ (left), $M(K^-\pi^+)$ and $M(\Sigma^0\bar K^{*0})$ (middle), and $M(K^0_S\pi^-)$ and $M(\Sigma^+K^{*-})$ (right) distributions from data samples. All components are indicated in the labels and described in the text.}\label{2dfit}
\end{figure*}

To extract the $\xic$ signal yields from two-body $\Lambda\bar K^{*0}/\Sigma^0\bar K^{*0}/\Sigma^+K^{*-}$ decay, we perform a two-dimensional (2D) binned maximum-likelihood fit to $M(K^-\pi^+/K^-\pi^+/K^0_S\pi^-)$ and $M(\Lambda\bar K^{*0}/\Sigma^0\bar K^{*0}/\Sigma^+K^{*-})$ distributions assuming there is no interference between the $\bar K^{*0}/\bar K^{*0}/K^{*-}$ signals and backgrounds. The 2D fitting function $f(M_1,M_2)$ is expressed as
\begin{equation}
f(M_1,M_2) = N^{\rm sig}s_1(M_1)s_2(M_2)+N^{\rm bg}_{\rm sb}s_1(M_1)b_2(M_2)+N^{\rm bg}_{\rm bs}b_1(M_1)s_2(M_2)+N^{\rm bg}_{\rm bb}b_1(M_1)b_2(M_2),
\end{equation}
where $s_1(M_1)$ and $b_1(M_1)$ are the signal and background probability density functions (PDFs) for the $M(\Lambda\bar K^{*0}/\Sigma^0\bar K^{*0}/\Sigma^+K^{*-})$ distributions, respectively, and $s_2(M_2)$ and $b_2(M_2)$ are the corresponding PDFs for the $M(K^-\pi^+/K^-\pi^+/K^0_S\pi^-)$ distributions. Here, $N^{\rm bg}_{\rm sb}$ and $N^{\rm bg}_{\rm bs}$ denote the numbers of peaking background events in $M(\Lambda\bar K^{*0}/\Sigma^0\bar K^{*0}/\Sigma^+K^{*-})$ and $M(K^-\pi^+/K^-\pi^+/K^0_S\pi^-)$ distributions, and $N^{\rm bg}_{\rm bb}$ is the number of combinatorial background in both $\xic$ and $\bar K^{*0}/K^{*-}$ candidates.
The signal shapes of $\xic$ and $\bar K^{*0}/K^{*-}$ ($s_1(M_1)$ and $s_2(M_2)$) are described by Breit-Wigner (BW) functions convolved with Gaussian functions, and second- or third-order polynomial functions represent the backgrounds ($b_1(M_1)$ and $b_2(M_2)$). The values of signal PDF parameters are fixed to those obtained from the fits to the corresponding simulated signal distributions. The values of the background shape parameters are free. The fitted results are shown in Fig.~\ref{2dfit}. The fitted $\xic$ signal yields in $\Xi^0_c\to \Lambda\bar K^{*0}$, $\Xi^0_c\to \Sigma^0\bar K^{*0}$, and $\Xi^0_c\to \Sigma^+K^{*-}$ are $3974\pm367$, $6260\pm254$, and $373\pm61$ events, respectively.

We calculate the branching fraction ratios according to the equations
\begin{equation} \label{eq:add1}
\frac{\BR(\xic\to\Lambda {\bar K}^{*0})}{\BR(\xic\to\Xi^-\pi^+)} = \frac{N^{\rm sig}_{\Lambda {\bar K}^{*0}}\cdot\varepsilon_{\Xi^-\pi^+}\cdot\BR(\Xi^-\to \Lambda\pi^-)}{N^{\rm sig}_{\Xi^-\pi^+}\cdot\varepsilon_{\Lambda {\bar K}^{*0}}\cdot\BR({\bar K}^{*0}\to K^-\pi^+)},
\end{equation}
\begin{equation} \label{eq:add2}
\frac{\BR(\xic\to\Sigma^0 {\bar K}^{*0})}{\BR(\xic\to\Xi^-\pi^+)} = \frac{N^{\rm sig}_{\Sigma^0 {\bar K}^{*0}}\cdot\varepsilon_{\Xi^-\pi^+}\cdot\BR(\Xi^-\to \Lambda\pi^-)}{N^{\rm sig}_{\Xi^-\pi^+}\cdot\varepsilon_{\Sigma^0 {\bar K}^{*0}}\cdot\BR(\Sigma^0\to \gamma\Lambda)\cdot\BR({\bar K}^{*0}\to K^-\pi^+)},
\end{equation}
and
\begin{equation} \label{eq:add3}
\begin{aligned}
\begin{split}
\frac{\BR(\xic\to\Sigma^+ K^{*-})}{\BR(\xic\to\Xi^-\pi^+)} = \frac{N^{\rm sig}_{\Sigma^+ K^{*-}}\cdot\varepsilon_{\Xi^-\pi^+}}{N^{\rm sig}_{\Xi^-\pi^+}\cdot\varepsilon_{\Sigma^+ K^{*-}}}\times~~~~~~~~~~~~~~~~~~~~~~~~~~~~~~~~~~~~~~~~\\
\frac{\BR(\Lambda\to p\pi^-)}{\BR(\Sigma^+\to p\pi^0)\cdot\BR(\pi^0\to\gamma\gamma)\cdot\BR(K^{*-}\to K^0_S\pi^-)\cdot\BR(K^0_S\to\pi^+\pi^-)}.
\end{split}
\end{aligned}
\end{equation}
\normalsize
Here, $N^{\rm sig}_{\Lambda {\bar K}^{*0}}$, $N^{\rm sig}_{\Sigma^0 {\bar K}^{*0}}$, $N^{\rm sig}_{\Sigma^+ K^{*-}}$, and $N^{\rm sig}_{\Xi^-\pi^+}$
are the signal yields, and $\varepsilon_{\Lambda {\bar K}^{*0}}$ = 18.9\%, $\varepsilon_{\Sigma^0 {\bar K}^{*0}}$ = 8.0\%, $\varepsilon_{\Sigma^+ K^{*-}}$ = 3.5\%, and $\varepsilon_{\Xi^-\pi^+}$ = 27.9\% are reconstruction efficiencies found from the signal simulations.
The reconstruction efficiencies do not include the branching fractions of intermediate states. In Eqs.~(\ref{eq:add1}) and (\ref{eq:add2}), the same intermediate branching fraction of $\BR(\Lambda\to p\pi^-)$ is canceled since the $\Lambda$ candidates are all reconstructed by $p\pi^-$.
Branching fractions $\BR(\Xi^-\to \Lambda\pi^-)$ = (99.887$\pm$0.035)\%, $\BR(\Sigma^0\to \gamma\Lambda)$ = 100\%, $\BR({\bar K}^{*0}\to K^-\pi^+)$ = 66.7\%, $\BR(\Lambda\to p\pi^{-})$ = (63.9$\pm$0.5)\%, $\BR(\Sigma^+\to p\pi^0)$ = (51.57$\pm$0.30)\%, $\BR(\pi^0\to\gamma\gamma)$ = (98.823$\pm$0.034)\%, $\BR(K^{*-}\to K^0_S\pi^-)$ = 33.34\%, and $\BR(K^0_S\to\pi^+\pi^-)$ = (69.2$\pm$0.05)\% are from PDG~\cite{PDG} directly or calculated based on the isospin symmetry. The branching fraction of $\xic \to \Xi^-\pi^+$ is (1.80$\pm$0.52)\%~\cite{Li}. Using the values above, we measure relative branching ratios to the normalization mode of $\Xi^0_c\to\Xi^-\pi^+$ and the branching fractions of $\Xi^0_c\to \Lambda\bar K^{*0}$, $\Xi^0_c\to \Sigma^0\bar K^{*0}$, and $\Xi^0_c\to \Sigma^+K^{*-}$, which are summarized in Table~\ref{br1}.

\begin{table}[htbp!]
\caption{The branching fractions and ratios, where the uncertainties are statistical, systematic, and from $\BR(\xic \to \Xi^-\pi^+)$~\cite{Li}.}\label{br1}
\vspace{0.2cm}
\centering
\begin{tabular}{c c}
\hline\hline
$\BR(\Xi^0_c\to \Lambda\bar K^{*0})/\BR(\xic\to \Xi^-\pi^+)$ & $0.18\pm0.02({\rm stat.})\pm0.01({\rm syst.})$ \\
$\BR(\Xi^0_c\to \Lambda\bar K^{*0})$ & $(3.3\pm0.3({\rm stat.})\pm0.2({\rm syst.})\pm1.0({\rm ref.}))\times10^{-3}$ \\\hline
$\BR(\Xi^0_c\to \Sigma^0\bar K^{*0})/\BR(\xic\to \Xi^-\pi^+)$ & $0.69\pm0.03({\rm stat.})\pm0.03({\rm syst.})$ \\
$\BR(\Xi^0_c\to \Sigma^0\bar K^{*0})$ & $(12.4\pm0.5({\rm stat.})\pm0.5({\rm syst.})\pm3.6({\rm ref.}))\times10^{-3}$ \\\hline
$\BR(\Xi^0_c\to \Sigma^+K^{*-})/\BR(\xic\to \Xi^-\pi^+)$ & $0.34\pm0.06({\rm stat.})\pm0.02({\rm syst.})$ \\
$\BR(\Xi^0_c\to \Sigma^+K^{*-})$ & $(6.1\pm1.0({\rm stat.})\pm0.4({\rm syst.})\pm1.8({\rm ref.}))\times10^{-3}$ \\
\hline\hline
\end{tabular}
\end{table}

\section{Asymmetry parameter extraction}

For $\xic\to\Lambda {\bar K}^{*0}$, the differential decay rate~\cite{053002} is given by:
\begin{equation} \label{eq:a1}
\frac{dN}{d{\rm cos}\theta_{\Lambda}}\propto 1+\alpha(\xic\to\Lambda {\bar K}^{*0})\alpha(\Lambda\to p\pi^-){\rm cos}\theta_{\Lambda},
\end{equation}
where $\alpha(\xic\to\Lambda {\bar K}^{*0})$ and $\alpha(\Lambda\to p\pi^-)$ are the asymmetry parameters of $\xic\to\Lambda {\bar K}^{*0}$ and $\Lambda\to p\pi^-$, and $\theta_\Lambda$ is the angle between the proton momentum vector and the opposite of $\xic$ momentum vector in the $\Lambda$ rest frame.

For $\xic\to\Sigma^0 {\bar K}^{*0}$, the differential decay rate~\cite{053002} is shown as:
\begin{equation} \label{eq:a2}
\frac{dN}{dcos\theta_{\Sigma^0}}\propto 1+\alpha(\xic\to\Sigma^0 {\bar K}^{*0})\alpha(\Sigma^0\to \gamma \Lambda){\rm cos}\theta_{\Sigma^0},
\end{equation}
where $\alpha(\xic\to\Sigma^0 {\bar K}^{*0})$ and $\alpha(\Sigma^0\to \gamma \Lambda)$ are the asymmetry parameters for $\xic\to\Sigma^0 {\bar K}^{*0}$ and $\Sigma^0\to \gamma \Lambda$, and $\theta_{\Sigma^0}$ is the angle between the $\Lambda$ momentum vector and the opposite of $\xic$ momentum vector in the $\Sigma^0$ rest frame. The value of $\alpha(\Sigma^0\to \gamma \Lambda)$ should be zero due to the conservation of parity for an electromagnetic decay. Thus, the distribution of ${\rm cos}\theta_{\Sigma^0}$ is expected to be flat.

For $\xic\to\Sigma^+K^{*-}$, the differential decay rate~\cite{053002} can be described with:
\begin{equation} \label{eq:a3}
\frac{dN}{dcos\theta_{\Sigma^+}}\propto 1+\alpha(\xic\to\Sigma^+K^{*-})\alpha(\Sigma^+\to p\pi^0){\rm cos}\theta_{\Sigma^+},
\end{equation}
where $\alpha(\xic\to\Sigma^+K^{*-})$ and $\alpha(\Sigma^+\to p\pi^0)$ are asymmetry parameters for $\xic\to\Sigma^+K^{*-}$ and $\Sigma^+\to p\pi^0$, and $\theta_{\Sigma^+}$ is the angle between the $p$ momentum vector and the opposite of $\xic$ momentum vector in the $\Sigma^+$ rest frame.

\begin{figure*}[htbp]
\centering
\includegraphics[width=3.2cm,angle=-90]{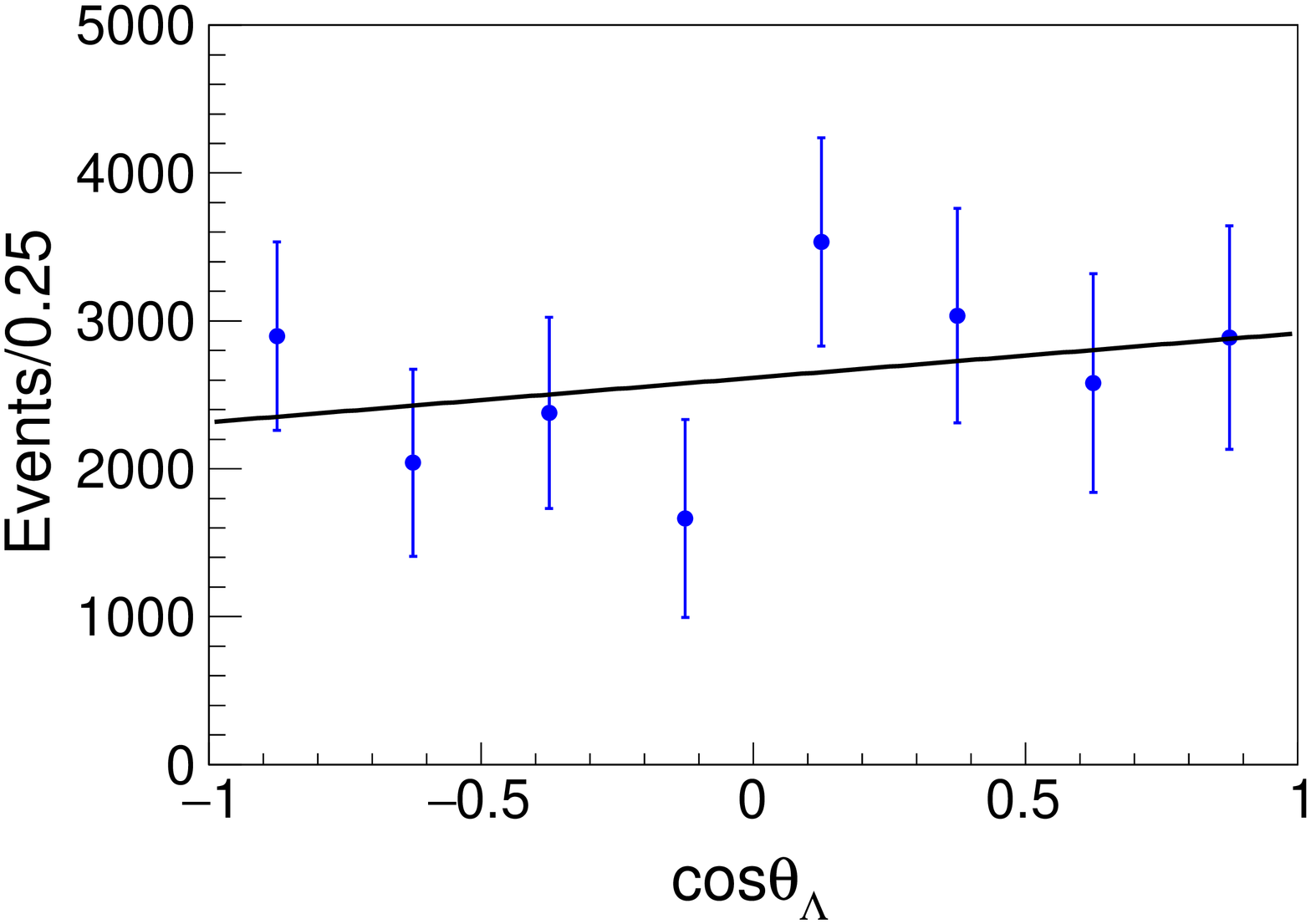}
\hspace{0.10cm}
\includegraphics[width=3.2cm,angle=-90]{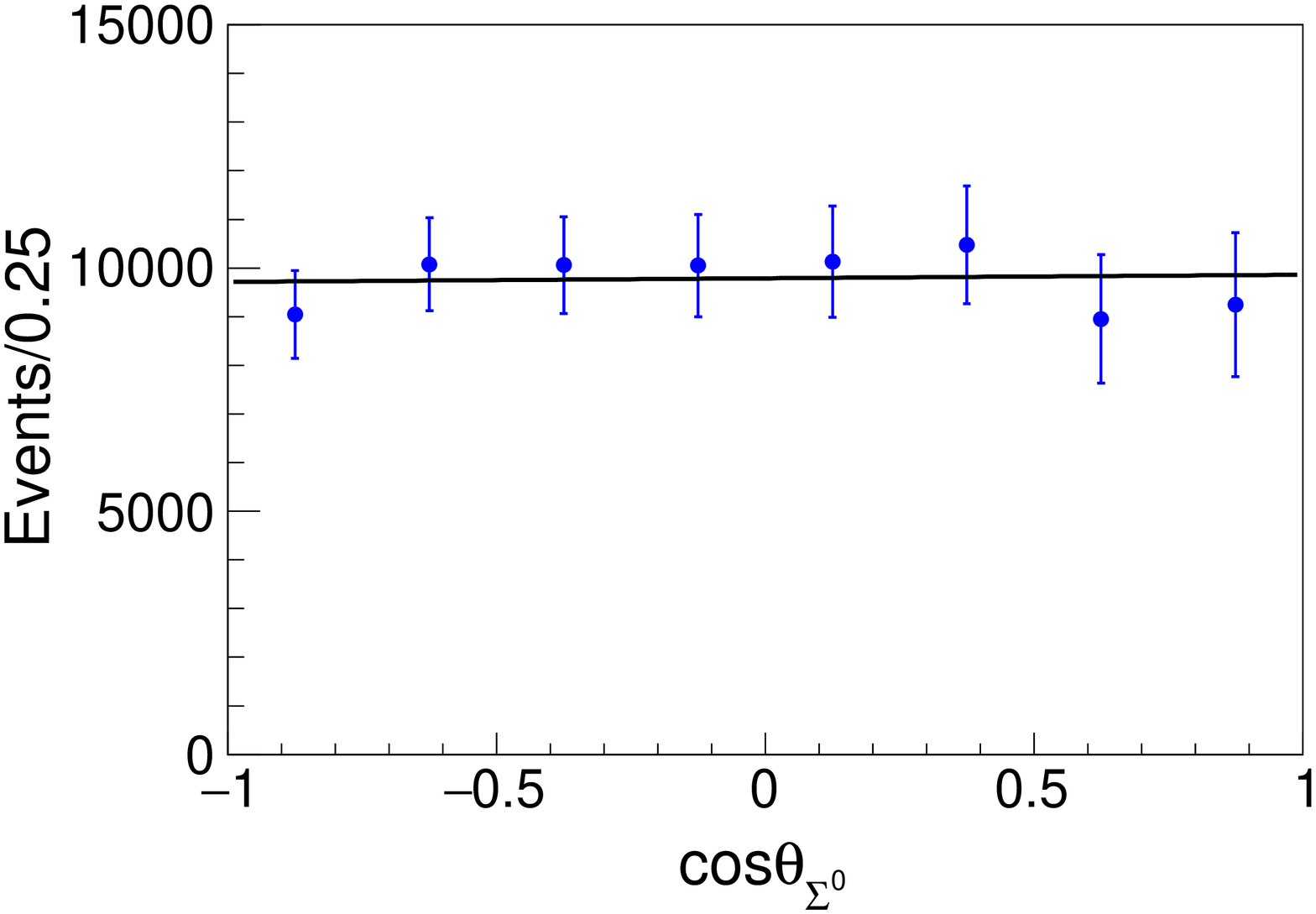}
\hspace{0.10cm}
\includegraphics[width=3.2cm,angle=-90]{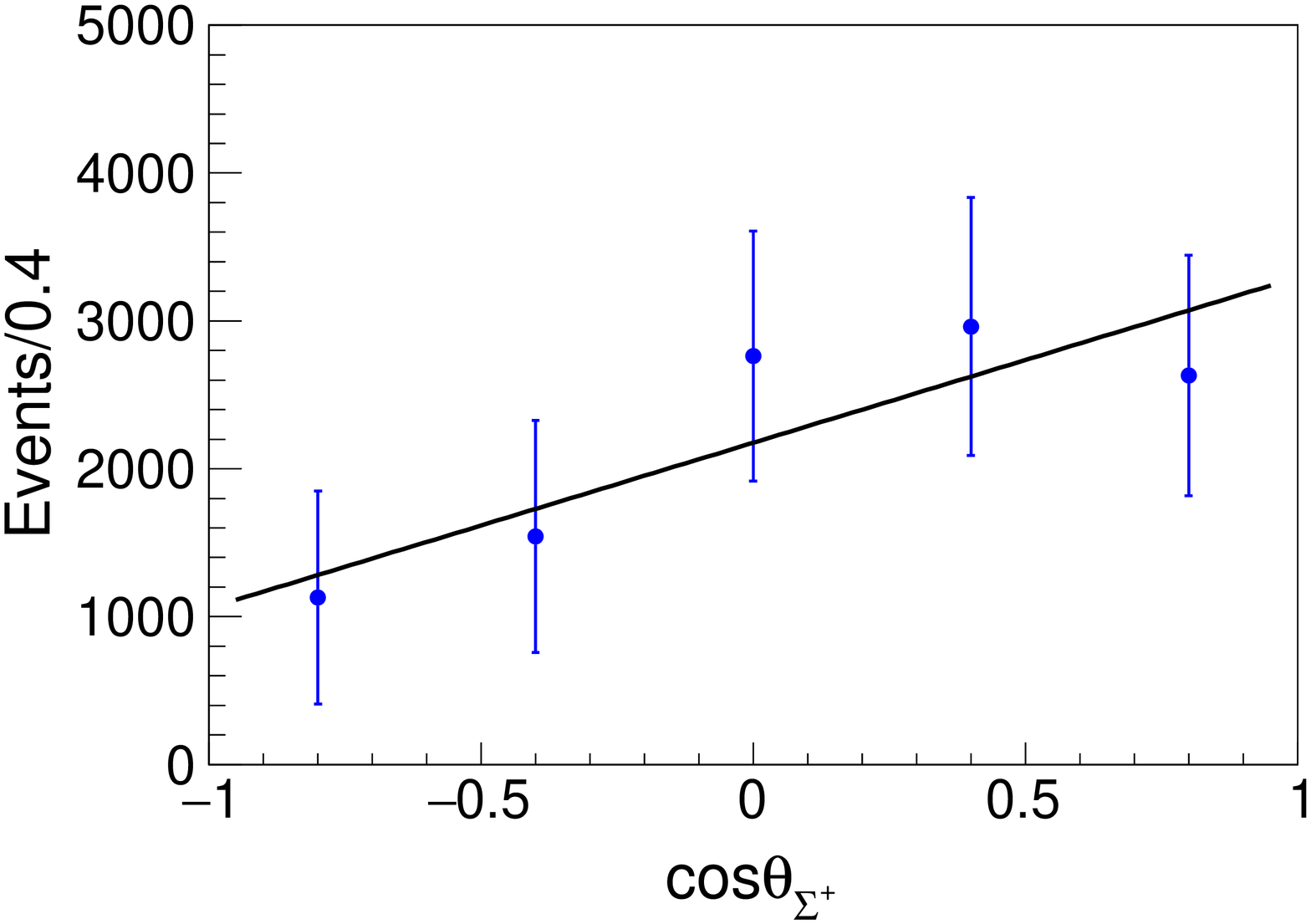}
\caption{The ${\rm cos}\theta_{\Lambda}$, ${\rm cos}\theta_{\Sigma^0}$, and ${\rm cos}\theta_{\Sigma^+}$ distributions after efficiency corrections from data samples. The lines show the fitted result with functions of Eq.~(\ref{eq:a1}), Eq.~(\ref{eq:a2}), and Eq.~(\ref{eq:a3}).}\label{asy}
\end{figure*}

We determine the asymmetry parameters by fitting the decay angular distributions of ${\rm cos}\theta_{\Lambda}$, ${\rm cos}\theta_{\Sigma^0}$, and ${\rm cos}\theta_{\Sigma^+}$ with Eqs.~(\ref{eq:a1})$-$(\ref{eq:a3}). For $\Xi^0_c\to \Lambda\bar K^{*0}$ and $\Xi^0_c\to \Sigma^0\bar K^{*0}$, we do 2D fits as above to data in 8 cos$\theta_{\Lambda}$ and cos$\theta_{\Sigma^0}$ bins. For $\Xi^0_c\to \Sigma^+K^{*-}$, we do 2D fits as above to data in 5 cos$\theta_{\Sigma^+}$ bins. The signal yields in data are summarized in Tables~\ref{tasy1} and \ref{tasy3}. We then make the efficiency-corrected ${\rm cos}\theta_{\Lambda}$, ${\rm cos}\theta_{\Sigma^0}$, and ${\rm cos}\theta_{\Sigma^+}$ distributions of data, as shown in Fig.~\ref{asy}. The fitted results are shown by the lines in Fig.~\ref{asy}. The returned values of product asymmetry parameters are listed in Table~\ref{asyt1}. The asymmetry parameter $\alpha(\Xi^0_c\to \Sigma^0\bar K^{*0})\alpha(\Sigma^0\to\gamma\Lambda)$ is $0.008\pm0.072({\rm stat.})\pm0.008({\rm syst.})$, which is consistent with zero, and compatible with parity conservation for an electromagnetic decay of $\Sigma^0\to\gamma\Lambda$.
Noting that $\alpha(\Lambda\to p\pi^-)$ = $0.747\pm0.010$ and $\alpha(\Sigma^+\to p\pi^0)$ = $-0.980\pm0.017$ from PDG~\cite{PDG}, the $\alpha(\Xi^0_c\to \Lambda\bar K^{*0})$ and $\alpha(\Xi^0_c\to \Sigma^+K^{*-})$ values are obtained and listed in Table~\ref{asyt1}. The systematic uncertainties are discussed below.

\begin{table}[htbp!]
\caption{The values of $\frac{\rm Signal~yield}{\rm Reconstruction~efficiency}$ from data samples in different cos$\theta_\Lambda$ and cos$\theta_{\Sigma^0}$ bins in $\Xi^0_c\to \Lambda\bar K^{*0}$ and $\Xi^0_c\to \Sigma^0\bar K^{*0}$.}\label{tasy1}
\small
\vspace{0.2cm}
\centering
\begin{tabular}{c c c c c c c c c}
\hline\hline
\centering
cos$\theta$ & $[-1, -0.75)$ & $[-0.75, -0.5)$ & $[-0.5, -0.25)$ & $[-0.25, 0)$ & $[0, 0.25)$ & $[0.25, 0.5)$ & $[0.5, 0.75)$ & $[0.75, 1]$ \\\hline
$\Lambda\bar K^{*0}$ & $\frac{582.1\pm128.2}{0.201}$ & $\frac{399.8\pm123.8}{0.196}$ & $\frac{456.1\pm124.0}{0.192}$ & $\frac{310.3\pm124.7}{0.187}$ & $\frac{644.3\pm128.4}{0.182}$ & $\frac{552.7\pm132.1}{0.182}$ & $\frac{477.0\pm136.8}{0.185}$ & $\frac{551.2\pm144.3}{0.191}$ \\\hline
$\Sigma^0\bar K^{*0}$ & $\frac{850.0\pm85.0}{0.094}$ & $\frac{905.8\pm86.0}{0.090}$ & $\frac{885.4\pm87.3}{0.088}$ & $\frac{837.9\pm87.4}{0.083}$ & $\frac{791.6\pm89.3}{0.078}$ & $\frac{782.7\pm90.5}{0.075}$ & $\frac{624.8\pm92.1}{0.070}$ & $\frac{581.5\pm93.3}{0.063}$ \\
\hline\hline
\end{tabular}
\end{table}

\begin{table}[htbp!]
\caption{The values of $\frac{\rm Signal~yield}{\rm Reconstruction~efficiency}$ from data samples in different cos$\theta_{\Sigma^+}$ bins in $\Xi^0_c\to \Sigma^+K^{*-}$.}\label{tasy3}
\vspace{0.2cm}
\centering
\begin{tabular}{c c c c c c}
\hline\hline
cos$\theta$ & $[-1, -0.6)$ & $[-0.6, -0.2)$ & $[-0.2, 0.2)$ & $[0.2, 0.6)$ & $[0.6, 1]$ \\\hline
$\Sigma^+K^{*-}$ & $\frac{44.1\pm28.1}{0.039}$ & $\frac{50.9\pm25.9}{0.033}$ & $\frac{88.4\pm27.1}{0.032}$ & $\frac{97.7\pm28.8}{0.033}$ & $\frac{92.1\pm28.4}{0.035}$ \\
\hline\hline
\end{tabular}
\end{table}

\begin{table}[htbp!]
\caption{The values of asymmetry parameters, where the uncertainties are statistical and systematic.}\label{asyt1}
\vspace{0.2cm}
\centering
\begin{tabular}{c | c}
\hline\hline
\centering
~~~$\alpha(\xic\to\Lambda {\bar K}^{*0})\alpha(\Lambda\to p\pi^-)$~~~ & ~~~$0.115\pm0.164({\rm stat.})\pm0.031({\rm syst.})$~~~ \\
$\alpha(\xic\to\Sigma^0 {\bar K}^{*0})\alpha(\Sigma^0\to \gamma \Lambda)$ & $0.008\pm0.072({\rm stat.})\pm0.008({\rm syst.})$ \\
$\alpha(\xic\to\Sigma^+K^{*-})\alpha(\Sigma^+\to p\pi^0)$ & $0.514\pm0.295({\rm stat.})\pm0.012({\rm syst.})$ \\\hline
$\alpha(\Xi^0_c\to \Lambda\bar K^{*0})$ & $0.15\pm0.22({\rm stat.})\pm0.04({\rm syst.})$ \\
$\alpha(\Xi^0_c\to \Sigma^+K^{*-})$ & $-0.52\pm0.30({\rm stat.})\pm0.02({\rm syst.})$ \\
\hline\hline
\end{tabular}
\end{table}

\section{Systematic Uncertainties}

The sources of systematic uncertainties for the branching fraction measurements include detection efficiency uncertainties, uncertainties in the  branching fractions of the intermediate states, and the uncertainty associated with the fitting technique used. Note that partial uncertainties from detection efficiency sources and branching fractions can be canceled by the reference mode of $\xic\to \Xi^-\pi^+$. The detection efficiency uncertainties include those from tracking efficiency (0.35\%/track), particle identification efficiency (1.1\%/kaon, 0.9\%/pion, and 2.9\%/proton), $K^0_S$ selection efficiency (2.23\%), $\pi^0$ reconstruction efficiency (2.25\%/$\pi^0$), and photon reconstruction efficiency (2.0\%/photon). The total detection efficiency uncertainties are obtained by adding all sources in quadrature.

As the partial uncertainties from branching fractions are canceled in the ratio to the reference mode, only the uncertainties of $\BR(\Sigma^+ \to p \pi^0)$ (0.6\%) and $\BR(\Lambda^0\to p\pi^-)$ (0.8\%)~\cite{PDG} need to be included for $\xic \to \Sigma^+K^{*-}$. In 2D fitting to $M(K^-\pi^+)$ and $M(\Lambda\bar K^{*0})$, $M(K^-\pi^+)$ and $M(\Sigma^0\bar K^{*0})$, and $M(K^0_S\pi^-)$ and $M(\Sigma^+K^{*-})$ distributions for $\Xi^0_c\to \Lambda\bar K^{*0}$, $\Xi^0_c\to \Sigma^0\bar K^{*0}$, and $\Xi^0_c\to \Sigma^+K^{*-}$, we enlarge the mass resolution by 10\%, and change the fit range and background shape, then the differences of signal yields are taken as the systematic uncertainties.
The total fit uncertainty is obtained by summing the uncertainties from $\Xi^0_c\to \Lambda\bar K^{*0}/\Sigma^0\bar K^{*0}/\Sigma^+K^{*-}$ and reference mode of $\Xi^0_c\to \Xi^-\pi^+$ in quadrature. All the uncertainties are summarized in Table~\ref{sys1}.
Finally, assuming all the sources are independent and adding them in quadrature, the total systematic uncertainties on the branching fraction measurements are calculated. The uncertainty of 28.9\% on $\BR(\Xi^0_c\to \Xi^-\pi^+)$~\cite{Li} is treated as systematic uncertainty separately.

\begin{table}[htbp!]
\caption{Relative systematic uncertainties (\%) in the branching fraction measurements.~The uncertainty of 28.9\% on $\BR(\Xi^0_c\to \Xi^-\pi^+)$~\cite{Li} is treated as an independent systematic uncertainty. }\label{sys1}
\vspace{0.2cm}
\centering
\begin{tabular}{c c c c}
\hline\hline
Final state & $\Lambda\bar K^{*0}$ & $\Sigma^0\bar K^{*0}$ & $\Sigma^+K^{*-}$ \\\hline
~~~Detection efficiency~~~ & 2.0 & 2.9 & 3.8 \\
Branching fraction & - & - & 1.0 \\
Fit uncertainty & 3.8 & 3.1 & 5.4 \\
Sum in quadrature & 4.3 & 4.2 & 6.7 \\
\hline\hline
\end{tabular}
\end{table}

The sources of the systematic uncertainties in the asymmetry parameter extractions include fitting procedures, the numbers of cos$\theta$ bins,  uncertainties in $\alpha(\Lambda\to p\pi^-)$ and $\alpha(\Sigma^+\to p\pi^0)$ values, and production polarization of $\xic$. The fitting uncertainties are estimated using simulated pseudoexperiments. We use an ensemble of simulated experiments to generate the mass spectra of $\Xi^0_c$ and ${\bar K}^*$ candidates corresponding to Fig.~\ref{2dfit}. The number of signal events in each ${\rm cos}\theta$ bin is obtained by a 2D fit to the generated mass spectra  after enlarging the mass resolution by 10\%, and changing the fit range and background shape. After 10,000 simulations, distributions of $\alpha(\xic\to\Lambda {\bar K}^{*0})\alpha(\Lambda\to p\pi^-)$, $\alpha(\xic\to\Sigma^0 {\bar K}^{*0})\alpha(\Sigma^0\to \gamma \Lambda)$, and $\alpha(\xic\to\Sigma^+K^{*-})\alpha(\Sigma^+\to p\pi^0)$ are obtained by fitting the slopes of the ${\rm cos}\theta_{\Lambda}$, ${\rm cos}\theta_{\Sigma^0}$ and ${\rm cos}\theta_{\Sigma^+}$ distributions. The differences between the fitted peaking values of the distributions of $\alpha(\xic\to\Lambda {\bar K}^{*0})\alpha(\Lambda\to p\pi^-)$, $\alpha(\xic\to\Sigma^0 {\bar K}^{*0})\alpha(\Sigma^0\to \gamma \Lambda)$, and $\alpha(\xic\to\Sigma^+K^{*-})\alpha(\Sigma^+\to p\pi^0)$ and the nominal values are taken as fitting uncertainties.

We change the numbers of cos$\theta$ bins from 8 and 5 to 10 and 8 for $\xic\to\Lambda {\bar K}^{*0},\Sigma^0 {\bar K}^{*0}$ and $\xic\to\Sigma^+K^{*-}$, and the differences on the asymmetry parameters are taken as the related systematic uncertainty.
The absolute uncertainties on $\alpha(\Lambda\to p\pi^-)$ and $\alpha(\Sigma^+\to p\pi^0)$ values are 0.002 and 0.009~\cite{PDG}.
This measurement is insensitive to production polarization of $\xic$, and no systematic error has been included from this source~\cite{111102}.
Finally, the absolute systematical uncertainties for $\alpha(\xic\to\Lambda {\bar K}^{*0})$, $\alpha(\xic\to\Sigma^0 {\bar K}^{*0})\alpha(\Sigma^0\to \gamma \Lambda)$, and $\alpha(\xic\to\Sigma^+K^{*-})$ are estimated by adding all individual uncertainties in quadrature. All the uncertainties are summarized in Table~\ref{sys2}.

\begin{table}[htbp!]
\caption{Absolute systematic uncertainties in the asymmetry parameter extractions.}\label{sys2}
\vspace{0.2cm}
\centering
\begin{tabular}{c c c c}
\hline\hline
Final state & $\Lambda\bar K^{*0}$ & $\Sigma^0\bar K^{*0}$ & $\Sigma^+K^{*-}$ \\\hline
Fit uncertainty & 0.023 & 0.006 & 0.009 \\
cos$\theta$ bins & 0.033 & 0.005 & 0.007 \\
~~~$\alpha(\Lambda\to p\pi^-)$ and $\alpha(\Sigma^+\to p\pi^0)$ values~~~ & 0.002 & $-$ & 0.009 \\\hline
Sum in quadrature & 0.041 & 0.008 & 0.015\\
\hline\hline
\end{tabular}
\end{table}

\section{Summary}

We measure for the first time the branching fractions and asymmetry parameters of $\Xi^0_c\to \Lambda\bar K^{*0}$, $\Xi^0_c\to \Sigma^0\bar K^{*0}$, and $\Xi^0_c\to \Sigma^+K^{*-}$ decays. The relative branching ratios to the normalization mode of $\Xi^0_c\to\Xi^-\pi^+$ and the branching fractions of $\Xi^0_c\to \Lambda\bar K^{*0}$, $\Xi^0_c\to \Sigma^0\bar K^{*0}$, and $\Xi^0_c\to \Sigma^+K^{*-}$ are calculated, as listed in Table~\ref{br1}.
We note that the branching fraction of $\Xi^0_c\to \Sigma^0\bar K^{*0}$ is much larger than that of $\Xi^0_c\to \Lambda\bar K^{*0}$, and this contradicts all the predictions based on $SU(3)_F$ flavor symmetry and dynamical models~\cite{659,5787,35,053002}.
This indicates the fraction of $\Lambda K^-\pi^+$ resonating through $\Lambda\bar K^{*0}$ is smaller than the fraction of $\Sigma^0 K^-\pi^+$ resonating through  $\Sigma^0 \bar K^{*0}$. The asymmetry parameters $\alpha(\Xi^0_c\to \Lambda\bar K^{*0})$ and $\alpha(\Xi^0_c\to \Sigma^+K^{*-})$ are measured to be $0.15\pm0.22({\rm stat.})\pm0.04({\rm syst.})$ and $-0.52\pm0.30({\rm stat.})\pm0.02({\rm syst.})$ with large statistical uncertainties.

\acknowledgments
We would like to thank Prof.~Chao-Qiang Geng for fruitful discussions. We thank the KEKB group for the excellent operation of the
accelerator; the KEK cryogenics group for the efficient
operation of the solenoid; and the KEK computer group, and the Pacific Northwest National
Laboratory (PNNL) Environmental Molecular Sciences Laboratory (EMSL)
computing group for strong computing support; and the National
Institute of Informatics, and Science Information NETwork 5 (SINET5) for
valuable network support.  We acknowledge support from
the Ministry of Education, Culture, Sports, Science, and
Technology (MEXT) of Japan, the Japan Society for the
Promotion of Science (JSPS), and the Tau-Lepton Physics
Research Center of Nagoya University;
the Australian Research Council including grants
DP180102629, 
DP170102389, 
DP170102204, 
DP150103061, 
FT130100303; 
Austrian Federal Ministry of Education, Science and Research (FWF) and
FWF Austrian Science Fund No.~P~31361-N36;
the National Natural Science Foundation of China under Contracts
No.~11435013,  
No.~11475187,  
No.~11521505,  
No.~11575017,  
No.~11675166,  
No.~11761141009;
No.~11975076;
No.~12042509;
Key Research Program of Frontier Sciences, Chinese Academy of Sciences (CAS), Grant No.~QYZDJ-SSW-SLH011; 
the  CAS Center for Excellence in Particle Physics (CCEPP); 
the Shanghai Pujiang Program under Grant No.~18PJ1401000;  
the Shanghai Science and Technology Committee (STCSM) under Grant No.~19ZR1403000; 
the Ministry of Education, Youth and Sports of the Czech
Republic under Contract No.~LTT17020;
Horizon 2020 ERC Advanced Grant No.~884719 and ERC Starting Grant No.~947006 ``InterLeptons'' (European Union);
the Carl Zeiss Foundation, the Deutsche Forschungsgemeinschaft, the
Excellence Cluster Universe, and the VolkswagenStiftung;
the Department of Atomic Energy (Project Identification No. RTI 4002) and the Department of Science and Technology of India;
the Istituto Nazionale di Fisica Nucleare of Italy;
National Research Foundation (NRF) of Korea Grant
Nos.~2016R1\-D1A1B\-01010135, 2016R1\-D1A1B\-02012900, 2018R1\-A2B\-3003643,
2018R1\-A6A1A\-06024970, 2018R1\-D1A1B\-07047294, 2019K1\-A3A7A\-09033840,
2019R1\-I1A3A\-01058933;
Radiation Science Research Institute, Foreign Large-size Research Facility Application Supporting project, the Global Science Experimental Data Hub Center of the Korea Institute of Science and Technology Information and KREONET/GLORIAD;
the Polish Ministry of Science and Higher Education and
the National Science Center;
the Ministry of Science and Higher Education of the Russian Federation, Agreement 14.W03.31.0026, 
and the HSE University Basic Research Program, Moscow; 
University of Tabuk research grants
S-1440-0321, S-0256-1438, and S-0280-1439 (Saudi Arabia);
the Slovenian Research Agency Grant Nos. J1-9124 and P1-0135;
Ikerbasque, Basque Foundation for Science, Spain;
the Swiss National Science Foundation;
the Ministry of Education and the Ministry of Science and Technology of Taiwan;
and the United States Department of Energy and the National Science Foundation.

\renewcommand{\baselinestretch}{1.2}


\begin{thebibliography}{**}
\addtolength{\itemsep}{-0.3 ex}

\bibitem{PDG} Particle Data Group, {\em Review of Particle Physics}, Prog. Theor. Exp. Phys. {\bf 2020} (2020) 083C01.
\bibitem{lifetime} LHCb Collaboration, {\em Precision measurement of the $\Lambda^+_c$, $\Sigma^+_c$ and $\Sigma^0_c$ baryon lifetimes}, Phys. Rev. D {\bf 100} (2019) 032001.
\bibitem{Li} Belle Collaboration, {\em First measurements of absolute branching fractions of the $\Xi^0_c$ baryon at Belle}, Phys. Rev. Lett. {\bf 122} (2019) 082001.
\bibitem{071101} LHCb Collaboration, {\em First branching fraction measurement of the suppressed decay $\Xi^0_c\to \pi^-\Lambda^+_c$}, Phys. Rev. D {\bf 102} (2020) 071101.
\bibitem{McNeil} Belle Collaboration, {\em Measurement of the resonant and non-resonant branching ratios in $\Xi^0_c\to \Xi^-K^+K^-$} (2020) arXiv:2012.05607.
\bibitem{1548} Belle Collaboration, {\em Measurements of the branching fractions of semileptonic decays $\Xi^0_c\to \Xi^-\ell^+\nu_{\ell}$ and asymmetry parameter ${\cal A}_{cp}$ of $\Xi^0_c\to \Xi^-\pi^+$ decay}, arXiv:2103.06496 (2021).

\bibitem{053002} C. Q. Geng, C. W. Liu, and T. H. Tsai, {\em Charmed baryon weak decays with vector mesons}, Phys. Rev. D {\bf 101} (2020) 053002.

\bibitem{00876} C. P. Jia, D. Wang, and F. S. Yu, {\em Charmed baryon decays in $SU(3)_F$ symmetry}, Nucl. Phys. B {\bf 956} (2020) 115048.
\bibitem{1527} M. J. Savage and R. P. Springer, {\em SU(3) predictions for charmed-baryon decays}, Phys. Rev. D {\bf 42} (1990) 1527.
\bibitem{414} M. J. Savage, {\em SU(3) violations in the nonleptonic decay of charmed hadrons}, Phys. Lett. B {\bf 257} (1991) 414.
\bibitem{7067} K. K. Sharma and R. C. Verma, {\em $SU(3)_{flavor}$ analysis of two-body weak decays of charmed baryons}, Phys. Rev. D {\bf 55} (1997) 7067.
\bibitem{2132} L. L. Chau, H. Y. Cheng, and B. Tseng, {\em Analysis of two-body decays of charmed baryons using the quark-diagram scheme}, Phys. Rev. D {\bf 54} (1996) 2132.
\bibitem{147} C. Q. Geng, Y. K. Hsiao, C. W. Liu, and T. H. Tsai, {\em Charmed baryon weak decays with SU(3) flavor symmetry}, J. High Energy Phys. {\bf 11} (2017) 147.
\bibitem{073006} C. Q. Geng, Y. K. Hsiao, C. W. Liu, and T. H. Tsai, {\em Anti-triplet charmed baryon decays with SU(3) flavor symmetry}, Phys. Rev. D {\bf 97} (2018) 073006.
\bibitem{593} C. Q. Geng, Y. K. Hsiao, C. W. Liu, and T. H. Tsai, {\em SU(3) symmetry breaking in charmed baryon decays}, Eur. Phys. J. C {\bf 78} (2018) 593.
\bibitem{073003} C. Q. Geng, Y. K. Hsiao, C. W. Liu, and T. H. Tsai, {\em Three-body charmed baryon decays with SU(3) flavor symmetry}, Phys. Rev. D {\bf 99} (2019) 073003.
\bibitem{19} C. Q. Geng, C. W. Liu, and T. H. Tsai, {\em Asymmetries of anti-triplet charmed baryon decays}, Phys. Lett. B {\bf 794} (2019) 19.
\bibitem{214} C. Q. Geng, C. W. Liu, T. H. Tsai, and S. W. Yeh, {\em Semileptonic decays of anti-triplet charmed baryons}, Phys. Lett. B {\bf 792} (2019) 214.
\bibitem{114022} C. Q. Geng, C. W. Liu, T. H. Tsai, and Y. Yu, {\em Charmed baryon weak decays with decuplet baryon and SU(3) flavor symmetry}, Phys. Rev. D {\bf 99} (2019) 114022.
\bibitem{946} J. Y. Cen, C. Q. Geng, C. W. Liu, and T. H. Tsai, {\em Up-down asymmetries of charmed baryon three-body decays}, Eur. Phys. J. C {\bf 79} (2019) 946.
\bibitem{165} H. J. Zhao, Y. K. Hsiao, and Y. Yao, {\em A diagrammatic analysis of two-body charmed baryon decays with flavor symmetry}, J. High Energy Phys. {\bf 02} (2020) 165.
\bibitem{066} D. Wang, P. F. Guo, W. H. Long, and F. S. Yu, {\em $K^0_S-K^0_L$ asymmetries and CP Violation in charmed baryon decays into neutral kaons}, J. High Energy Phys. {\bf 03} (2018) 066.
\bibitem{03480} X. G. He, Y. J. Shi, and W. Wang, {\em Uni?cation of flavor SU(3) analyses of heavy hadron weak decays}, Eur. Phys. J. C {\bf 80} (2020) 359.
\bibitem{35} Y. K. Hsiao, Y. Yao, and H. J. Zhao, {\em Two-body charmed baryon decays involving vector meson with SU(3) flavor symmetry}, Phys. Lett. B {\bf 792} (2019) 35.
\bibitem{036018} S. Roy, R. Sinha, and N. G. Deshpande, {\em Non-leptonic beauty baryon decays and CP-asymmetries based on SU(3)-flavor analysis}, Phys. Rev. D {\bf 101} (2020) 036018.
\bibitem{659} J. G. K\"orner and M. Kr\"amer, {\em Exclusive non-leptonic charm baryon decays}, Z. Phys. C {\bf 55} (1992) 659.
\bibitem{3836} Q. P. Xu and A. N. Kamal, {\em Nonleptonic charmed-baryon decays: $B_c\to$ B(3/2$^{+}$, decuplet) + P(0$^-$) or V(1$^-$)}, Phys. Rev. D {\bf 46} (1992) 3836.
\bibitem{270} Q. P. Xu and A. N. Kamal, {\em Cabibbo-favored nonleptonic decays of charmed baryons}, Phys. Rev. D {\bf 46} (1992) 270.
\bibitem{4188} H. Y. Cheng and B. Tseng, {\em Cabibbo-allowed nonleptonic weak decays of charmed baryons}, Phys. Rev. D {\bf 48} (1993) 4188.
\bibitem{014011} J. Zou, F. Xu, G. Meng, and H. Y. Cheng, {\em Two-body hadronic weak decays of antitriplet charmed baryons}, Phys. Rev. D {\bf 101} (2020) 014011.
\bibitem{5787} P. $\dot Z$enczykowski, {\em Nonleptonic charmed-baryon decays: Symmetry properties of parity-violating amplitudes}, Phys. Rev. D {\bf 50} (1994) 5787.
\bibitem{3417} T. Uppal, R. C. Verma, and M. P. Khanna, {\em Constituent quark model analysis of weak mesonic decays of charm baryons}, Phys. Rev. D {\bf 49} (1994) 3417.
\bibitem{5632} M. A. Ivanov, J. G. K\"orner, V. E. Lyubovitskij, and A. G. Rusetsky, {\em Exclusive nonleptonic decays of bottom and charm baryons in a relativistic three-quark model: evaluation of nonfactorizing diagrams}, Phys. Rev. D {\bf 57} (1998) 5632.
\bibitem{217} K. K. Sharma and R. C. Verma, {\em A study of weak mesonic decays of $\Lambda_c$ and $\Xi_c$ baryons on the basis of HQET results}, Eur. Phys. J. C {\bf 7} (1999) 217.
\bibitem{Belle1} Belle Collaboration, {\em The Belle Detector}, Nucl. Instr. and Methods Phys. Res. Sect. A {\bf 479} (2002) 117.
\bibitem{Belle2} Belle Collaboration, {\em Physics achievements from the Belle experiment}, Prog. Theor. Exp. Phys. {\bf 2012} (2012) 04D001.
\bibitem{KEKB1} S. Kurokawa and E. Kikutani, {\em Overview of the KEKB accelerators}, Nucl. Instr. and Methods Phys. Res. Sect. A {\bf 499} (2003) 1, and other papers included in this volume.
\bibitem{KEKB2} T. Abe et al., {\em Achievements of KEKB}, Prog. Theor. Exp. Phys. {\bf 2013} (2013) 03A001, and references therein.
\bibitem{EVTGEN} D. J. Lange, {\em The EvtGen particle decay simulation package}, Nucl. Instr. and Methods Phys. Res. Sect. A {\bf 462} (2001) 152.
\bibitem{PYTHIA} T. Sj\"ostrand, S. Mrenna and P. Skands, {\em PYTHIA 6.4 Physics and Manual}, J. High Energy Phys. {\bf 0605} (2006) 026.
\bibitem{291} E. Barberio and Z. W\c as, {\em PHOTOS: A universal Monte Carlo for QED radiative corrections: Version 2.0}, Comput. Phys. Commun. {\bf 79} (1994) 291.
\bibitem{geant3} R. Brun et al., {\em GEANT}, CERN Report No. DD/EE/84-1 (1984).
\bibitem{PID} E. Nakano, {\em Belle PID}, Nucl. Instr. and Methods Phys. Res. Sect. A {\bf 494} (2002) 402.
\bibitem{190} M. Feindt and U. Kerzel, {\em The NeuroBayes neural network package}, Nucl. Instr. and Methods Phys. Res. Sect. A {\bf 559} (2006) 190.
\bibitem{2014} H. Nakano, {\em Search for new physics by a time-dependent CP violation analysis of the decay $B\to K^0_S \eta \gamma$ using the Belle detector}, Ph.D Thesis, Tohoku University (2014) Chapter 4, {\url {http://hdl.handle.net/10097/58814}}.
\bibitem{052011} Belle Collaboration, {\em Study of excited $\Xi_c$ states decaying into $\Xi^0_c$ and $\Xi^+_c$ baryons}, Phys. Rev. D {\bf 94} (2016) 052011.
\bibitem{107540} X. Y. Zhou, S. X. Du, G. Li, and C. P. Shen, {\em TopoAna: A generic tool for the event type analysis of inclusive Monte-Carlo samples in high energy physics experiments}, Comput. Phys. Commun. {\bf 258} (2021) 107540.
\bibitem{111102} CLEO Collaboration, {\em Measurement of the decay asymmetry parameters in $\Xi^0_c\to \Xi^-\pi^+$}, Phys. Rev. D {\bf 63} (2001) 111102.

\end{thebibliography}
\end{document}